\documentclass[journal=jacsat,manuscript=article]{achemso}

\usepackage{chemformula} 
\usepackage[T1]{fontenc} 

\usepackage{graphicx}
\usepackage{color}
\definecolor{lightgrey}{rgb}{0.8627,0.8627,0.8627}
\usepackage{amsmath,amssymb}
\usepackage{booktabs}
\usepackage{colortbl}

\newcommand{\ds}{\displaystyle}

\def\<{\langle}
\def\>{\rangle}

\author{Timoteo Carletti}
\affiliation{naXys, Namur Institute for Complex Systems, 
             University of Namur, 
             8 rempart de la Vierge, B5000, Namur, Belgium}
\author{Duccio Fanelli}
\affiliation{Dipartimento di Fisica e Astronomia, Universit\`{a}  
             di Firenze, INFN and CSDC, 
             Via Sansone 1, 50019 Sesto Fiorentino, Firenze, Italy}
\author{Francesco Piazza}
\affiliation{Universit\'e d'Orl\'eans and Centre de Biophysique 
             Mol\'eculaire, CNRS-UPR4301,
             Rue C. Sadron, 45071, Orl\'eans, France}

\title[\small The unreasonable effectiveness of 
                simple models]
      {COVID-19: The unreasonable effectiveness of simple models}

\abbreviations{}

\begin{document}

%
\begin{abstract}
\noindent When the novel coronavirus disease SARS-CoV2 (COVID-19) 
was officially declared a pandemic by the WHO in March 2020, 
the scientific community had already braced up in the effort of making sense of the 
fast-growing wealth of data gathered by national authorities 
all over the world. However, despite the diversity of novel theoretical 
approaches and the comprehensiveness of many widely established 
models, the official figures that recount the course of the outbreak still 
sketch a largely elusive and intimidating picture.
Here we show unambiguously that the dynamics of the COVID-19 outbreak belongs 
to the simple universality class of the SIR model and extensions 
thereof.  Our analysis naturally leads us to establish that 
there exists a fundamental limitation to any theoretical approach,
namely the unpredictable non-stationarity of the testing frames behind
the reported figures. However, we show how such bias can be 
quantified self-consistently and employed to mine useful and accurate information 
from the data. In particular, we describe how the time evolution of the 
reporting rates controls the occurrence of the apparent epidemic peak,
which typically follows the true one in countries that were not 
vigorous enough in their testing at the onset of the outbreak.
The importance of testing early and resolutely appears as a natural corollary 
of our analysis, as countries that tested massively at the start 
clearly had their true peak earlier and less deaths overall.    
\end{abstract}

\section{Introduction}

In December 2019  the novel coronavirus disease SARS-CoV2 (COVID-19) 
emerged in Wuhan, China. Despite the drastic containment measures implemented by the Chinese government, 
the disease quickly  spread all over the world,
officially reaching pandemic level in March 2020~\cite{WHO:2020aa}. 
From the beginning of the outbreak, the international community braced up in the commendable effort
to rationalise  the massive load of data~\cite{Layne:2020aa,Enserink:2020aa}, 
painstakingly collected by local authorities throughout the world.  
However, epidemic spreading is a complex process, whose details depend
to a large extent on the behaviour of the virus~\cite{Wang:2020aa,Kupferschmidt:2020aa} and 
of its current principal vectors (we, the humans). 
Additionally, recorded data necessarily suffer from many kinds of bias, so that 
much information is intrinsically missing in the data released officially. 
For example, the total number of reported  infected bears the imprint 
of the growing size of the sampled population. This 
major source of non-stationarity should be properly gauged if one 
aims at explaining the observed trends and elaborate 
forecasts for the epidemics evolution. Similarly, the time series of recovered 
individuals that  are released officially are heavily unreliable, 
due to the uneven ability of different countries to trace 
asymptomatic or mildly symptomatic patients. \\
\indent In this letter, we  demonstrate that  overlooking these effects  inevitably 
leads to fundamentally wrong predictions, irrespective of whether mean-field, compartment-based 
models~\cite{Gatto:2020aa,Giordano:2020aa}  or more sophisticate individual-based, stochastic 
spatial epidemic models~\cite{Chinazzi:2020aa,Balcan:2009aa} are employed. More precisely, it is important to 
realise that the populations of infected reported by national agencies
do not provide a reliable picture to correctly identify the epidemics peaks.
The primary cause resides in the marked {\em non-stationary} character of 
the testing activity. However, while sources of bias due to 
differences in sampling and reporting policies among different countries 
are generally acknowledged, notably 
by studies applying Bayesian inference~\cite{Roda:2020aa,De-Brouwer:2020aa} 
and causal models~\cite{Hitchcock:2020aa},
surprisingly the critical issue of non-stationarity appears to have been significantly
overlooked in the literature on COVID-19. Nonetheless, 
as we show in the following, save few creditable exceptions, the 
reporting rate has generally increased substantially as the outbreak was gathering momentum, along with
the stepping up of the number of tests conducted. Let us stress this  point 
once more -- any modelling efforts will 
be pointless unless non-stationarity of reporting is properly accounted for. \\
\indent While infected and recovered bear the marks of the convolution with 
unpredictably non-stationary reading frames,  it can be argued that
the time series of deceased individuals provide a more reliable picture of the true, bias-free
underlying dynamics of the disease. In fact, while there might have been 
inaccuracies in attributing deaths to COVID-19, it can be surmised that these should lead to
systematic and reasonably stationary errors.  \\
\indent Contrary to intuition, and to an excellent degree of 
accuracy, a death-only analysis of the data shows that COVID-19 outbreaks 
in all countries fall in the simple SIR universality class,  
mean-field compartmental models originated by the the seminal 1927 paper by 
Kermack and McKendrick~\cite{Kermack:1927aa}. 
Importantly, this fact can only be brought to the fore in the reduced space of deaths and their 
time derivatives, as this allows one to filter out virtually all spurious effects associated with 
non-stationary reading frames, which have been systematically underestimated in the vast 
majority of reports. \\
\indent Taken together, our analysis suggests that simple mean-field models can be meaningfully 
invoked to gather a quantitative and robust picture of the COVID-19 epidemic spreading. 
Additionally, as a natural byproduct of our reasoning, we were able to calculate the time-dependent 
reporting fractions for infected and recovered individuals. These provide a clear a-posteriori reading 
frame for the relative position of the true epidemic peak and that observed from the reported data,
which is found to be typically shifted away artificially in the future.   
Furthermore, casting the SIRD dynamics as a function of the deaths time series and its derivative
leads us to derive simple formulae  that prove invaluable for real-time monitoring of the disease. 
As an important corollary of our analysis, doubts can be cast on all procedures 
targeted at estimating the reproduction index of the epidemics based on the {\em raw} time series of 
infected (and recovered), i.e. not corrected for non-stationary testing.

\section{How to look at the reported figures}

\indent Within a simple mean-field compartmental model, people are divided into different  
{\em species}~\cite{Fanelli:2020aa}. For example, in the susceptible (S), infected (I), recovered (R), 
dead (D) scheme (SIRD)~\cite{Anastassopoulou:2020aa}, any individual in the fraction of the 
overall population that will eventually get sick belongs to one of the aforementioned classes.  
The latter represent the different stages of the disease and their size is used to model its time course. 
Let $S_0 = S(0)$ be the size of the initial  pool of susceptible individuals. Then the SIRD dynamics is 
governed by the following set of coupled differential equations
\begin{equation}
\label{eq:SIRD}
\begin{cases}
\dot{S} &= -\frac{\ds r}{\ds S_0}SI\\
\dot{I} &= \frac{\ds r}{\ds S_0}SI-\left( a+d\right) I \\
\dot{R} &= a I\\
\dot{D} &= d I\, 
\end{cases}
\end{equation}
where $\dot{X}$ stand for the time derivative of $X=S,I,R,D$.
The parameter $r$ represents the infection rate, i.e. the probability per unit time that 
a susceptible individual contracts the disease when she enters in contact with an infected person. 
The parameters $a$ and $d$ denote the recovery and death rates, respectively. Note that,
$S(t)+I(t)+R(t)+D(t)=S_0+I_0$ is an invariant of the SIRD dynamics, i.e. it remains constant 
as time elapses~\footnote{It is reasonable to assume that $R(0)=0$ as this coronavirus 
has never been in contact with humans before. Furthermore, $S(0) \gg I(0) \simeq \mathcal{O}(1)$,
which justifies the way Eqs.~\eqref{eq:SIRD} are written down (see also Methods).}
In general, one could attempt to calibrate model~\eqref{eq:SIRD}
against the available time series for the infected, 
recovered and deaths, so as to determine the best-fit estimates of  
$r$, $a$, $d$ and $S_0$. However, it may be argued that this  procedure amounts to over-fitting,
due to the intrinsic high correlation between $r$ and $S_0$ in the model. More commonly, 
a face-value figure for $S_0$ is assumed from the start, e.g. the size of a country or of some 
specific region. However, this choice is somehow arbitrary, as the {\em true} basin of the epidemics 
can in principle be determined meaningfully only a posteriori.\\
\indent More importantly, as argued above,   
there is a more fundamental limitation to any  attempt of adjusting a model to the raw figures
beyond specific, model-dependent issues. The cause is the time-varying bias in the reported data. 
Essentially, this translates  into an a-priori unknown discrepancy between the true size of 
infected and recovered and the figures released officially. 
The key observation at this point is that the time series of deaths, $D(t)$, can be 
regarded as the only data set that returns a 
faithful representation of the epidemics course. More precisely, it appears reasonable to surmise that 
all unknown sources of errors on deaths can be considered as roughly stationary over the time span 
considered. \\
\indent It is well-known that the SIR dynamics can be recast in a particularly 
appealing form in the space $(\dot{R},R)$, $R$ denoting in this case the overall class of recovered and 
deceased~\cite{Murray:1989aa,Vattay:2020aa}. 
When one distinguishes between the two populations $D$ and $R$ (the recovered), 
as in the SIRD dynamics~\eqref{eq:SIRD}, a similar manipulation leads to (see Methods)
\begin{equation}
\label{eq:dotDD}
\dot{D} = f(D) \stackrel{\rm def}{=}  \gamma \left( 1-e^{-\xi D} \right) - \beta D
\end{equation}
where $\gamma=dS_0$, $\xi=r/(d S_0)$ and $\beta=a+d$. The differential equation~\eqref{eq:dotDD}
is equivalent to the full set of equations~\eqref{eq:SIRD} and the associated conservation law. 
%
\begin{figure}[t!]
\centering
\includegraphics[width=\textwidth]{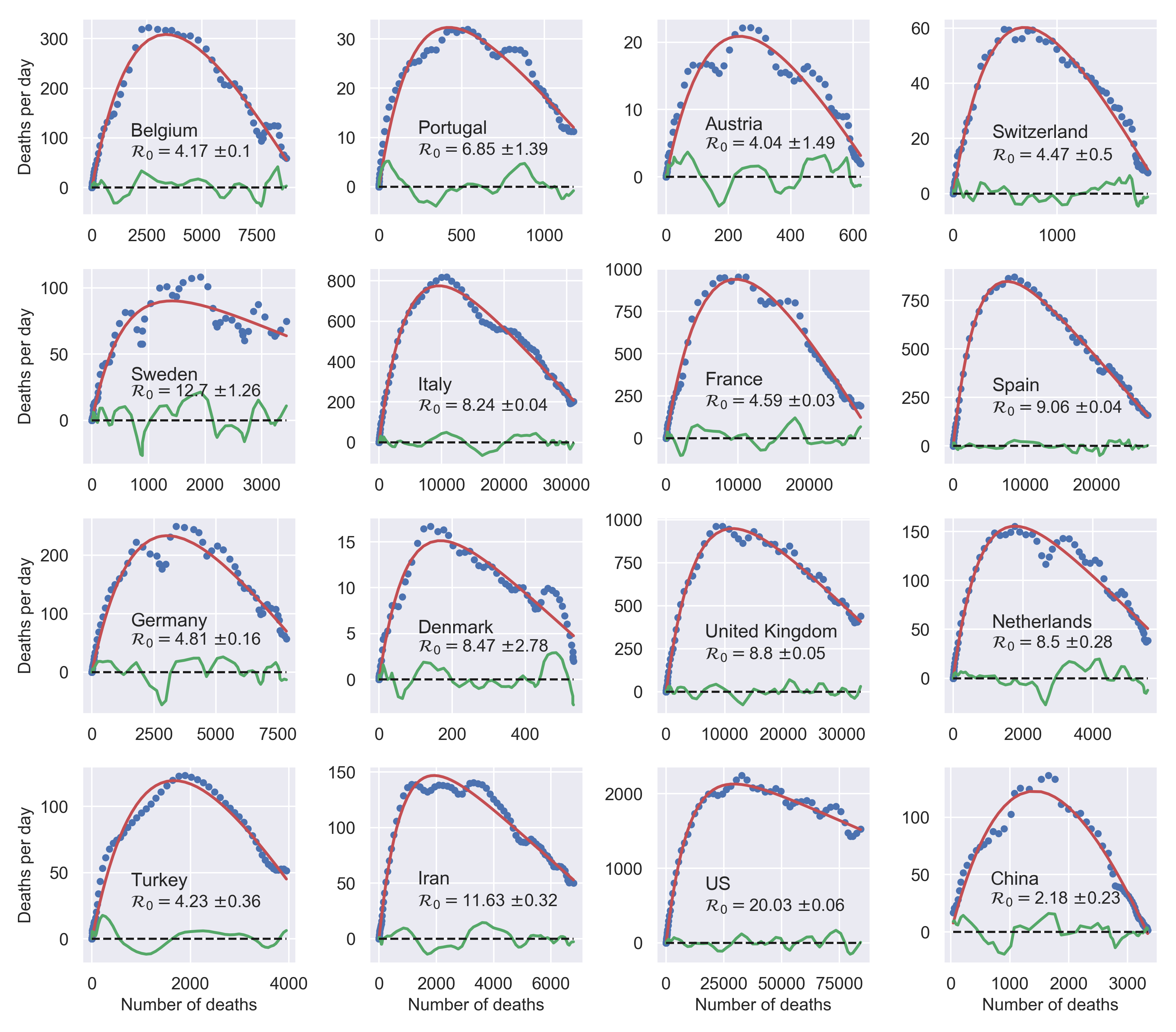}
\caption{\textbf{Universal behaviour of the COVID-19 outbreak in the $(\dot{D},D)$ plane}. 
When analysed in this space, outbreaks in different countries are described by the same 
three-parameters function $f(D)$, Eq.~\eqref{eq:dotDD}. The reported values of the effective 
reproduction number at $t=0$, $\mathcal{R}_0$ are computed as $\mathcal{R}_0 =\xi \gamma/\beta$. 
The data for $\dot{D}$ are centred rolling averages performed over a window of 6 days. 
%
%
Residuals most likely contain 
the (possibly time-shifted) blueprints of specific containment measures~\cite{Dehning:2020aa}.}
\label{fig:phiR0x}
\end{figure}
%
Overall, it provides an expedient and virtually bias-free means of testing whether 
the time evolution of an outbreak is governed by a SIRD-like dynamics. \\
\indent Fig.~\ref{fig:phiR0x} shows very clearly that the COVID-19 outbreak falls in the SIRD 
universality class. The data reported by national authorities in different countries are plotted in the 
plane ($\dot{D}$,$D$) and compared with three-parameters fits performed with 
Eq.~\eqref{eq:dotDD}. 
The quality of the fits appears remarkably good, irrespective of the country and the actual severity 
of the epidemics in terms of deaths and deaths per day. 
Residual oscillations about the mean-field interpolating trends are visible, 
pointing to interesting next-to-leading order corrections, which would deserve further attention. \\
\indent It appears remarkable that a process that is inherently complex and unpredictably 
many-faceted can be accurately described in terms of an effective, highly simplified mean-field scheme, 
where many factors are deliberately omitted, such as space 
and different sorts of population stratification. 
%
%
%
Of course, the effectiveness of SIR schemes and simple modifications thereof 
are well-known~\cite{Brauer:2012aa,Diekmann:2000aa}. 
For example, a recent work shows that a modified SIR model where the infection rate is let 
vary with time can be employed to infer change points in the epidemic spread to gauge 
the effectiveness of specific confinement measures in Germany~\cite{Dehning:2020aa}. 
In an earlier work, two of the present authors had considered a similar approach to 
elaborate predictions of the outbreak in Italy that proved fundamentally right when 
compared with the evolution observed afterwards~\cite{Fanelli:2020aa}.
What Fig.~\ref{fig:phiR0x} critically adds to the picture is a bias-free confirmation 
that simple mean-field schemes can be used meaningfully to draw quantitative conclusions. \\
\indent The best-fit values of the reproduction number $\mathcal{R}_0$ extracted from the above analysis
should be considered as {\em effective} estimates of the overall severity of the outbreak in 
different countries.  In particular, it can be argued that these should incorporate all 
sources of non-stationarity left besides the time-dependent bias associated with the 
evolution of the testing rate. Most likely,  these amount to different degrees of (time-dependent) 
reduction of the infection rate $r$, following the containment measures. 
Interestingly, we find that, rather generally, the evolution of such non-stationary SIRD dynamics falls 
into the same universality class in the plane $(\dot{D},D)$ as signified by $f(D)$, 
Eq.~\eqref{eq:dotDD} (see also section 3 in the 
supplementary material). More precisely, it turns out that a reduction 
of $r$ by a factor $f<1$ over a certain time window simply corresponds to a proportional rescaling 
of the reproduction number to a lower value, $\mathcal{R}_0 \to \mathcal{R}_0 f$.
A similar conclusion holds if the non-stationarity causes the opposite effect, 
that is, $f>1$. This might be the case, for example, of overlooked hotbeds or 
large undetected gatherings that cause the infection to accelerate 
at a certain point in time.
Remarkably, whatever the direction, it can be proved that this rescaling 
is independent of the typical time scale over which
the transition $r \to fr$ unfolds, as well as the point in time that marks its onset
(see supplementary material).\\
%
%
\indent An important corollary of our analysis is that the observed epidemic peak
displayed by the time series of the {\em measured}
infected, $I_M(t)$, typically occurs  
later than the true peak, the more so the more the testing rate is ramped up 
during the outbreak. The correct occurrence of the peak and a comprehensive understanding 
of how the testing activity controls the apparent evolution of the epidemics can 
be obtained with the following simple argument. 
In general, one can define two time-dependent {\em reporting fractions}, 
gauging the unknown disparity between the true and the measured populations
\begin{equation}
\label{e:alphadef}
\alpha_I(t) = \frac{I_M(t)}{I(t)} \qquad 
\alpha_R(t) = \frac{R_M(t)}{R(t)}
\end{equation}
where $R_M(t)$ refers to the reported population of recovered persons.
Taking into account the first definition in Eqs.~\eqref{e:alphadef}, differentiation 
of the last equation in~\eqref{eq:SIRD} with respect to time gives
\begin{equation}
\label{e:ddotD}
\ddot{D}(t) = \frac{d}{\alpha_I(t)}\dot{I}_M(t) - 
\frac{d}{\alpha_I^2(t)}  I_M(t) \dot{\alpha}_I(t) 
\end{equation}
The apparent epidemic peak occurs when $\dot{I}_M(t)=0$, whereas we know from the fact that 
the epidemics is governed by a SIRD-like dynamics that the true peak occurs 
when $\ddot{D}=0$, i.e. when the number of deaths/day reaches a maximum. 
Thus, from Eq.~\eqref{e:ddotD}
one immediately sees that whether the apparent peak is observed before or after the {\em true} peak 
depends on the sign of the rate of reporting,
$\dot{\alpha}_I(t)$.  More precisely, if the testing activity is steadily ramping up $(\dot{\alpha}_I>0)$, 
the true peak will occur earlier than the apparent one. This is because 
the reported infected will have a maximum for  $\ddot{D}<0$, 
i.e. past the maximum of the $\dot{D}$. 
Conversely, if the testing rate decreases, this will anticipate the apparent peak, 
giving the false impression that the worst might be over while the true number 
of infected is in fact still increasing.  We find that this analysis applies to all countries 
considered in this paper (see also supplementary material), whereby either the former or
the latter scenarios are invariably observed. \\
%
\begin{figure}[t!]
\centering
\begin{tabular}{ccc}
Italy & Germany & China \\
\includegraphics [width=0.33\textwidth]{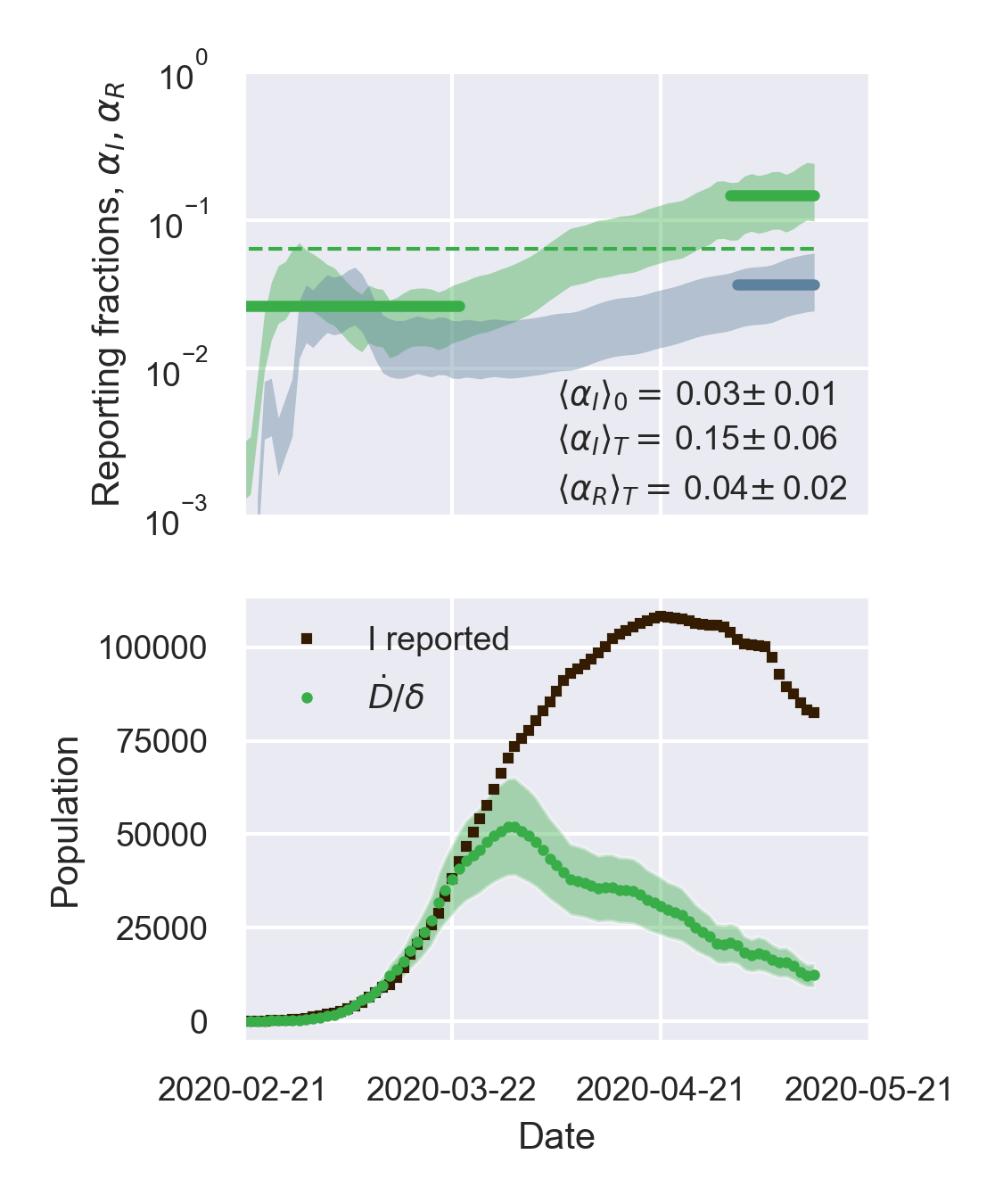} &  
\includegraphics [width=0.33\textwidth]{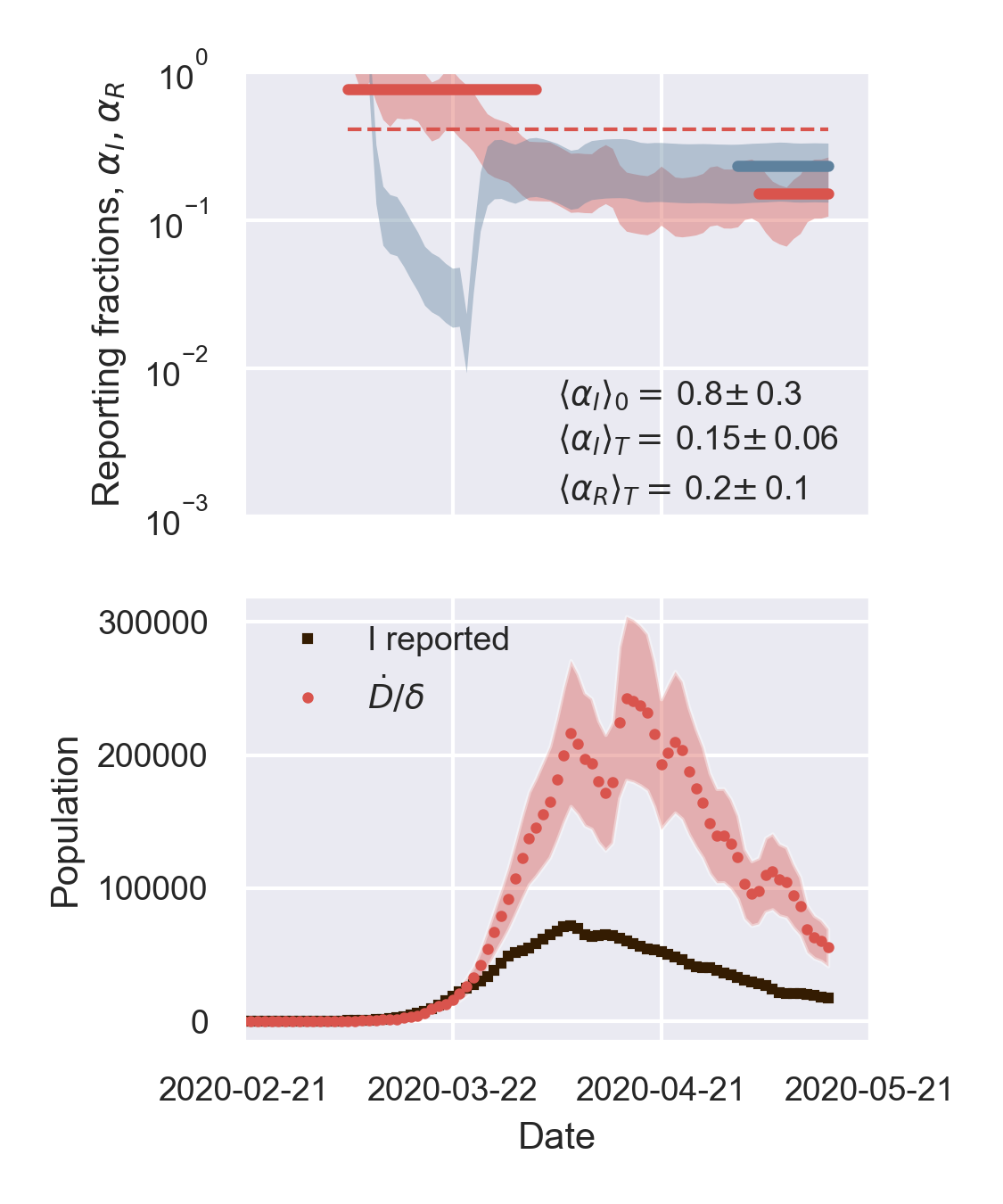} & 
\includegraphics [width=0.33\textwidth]{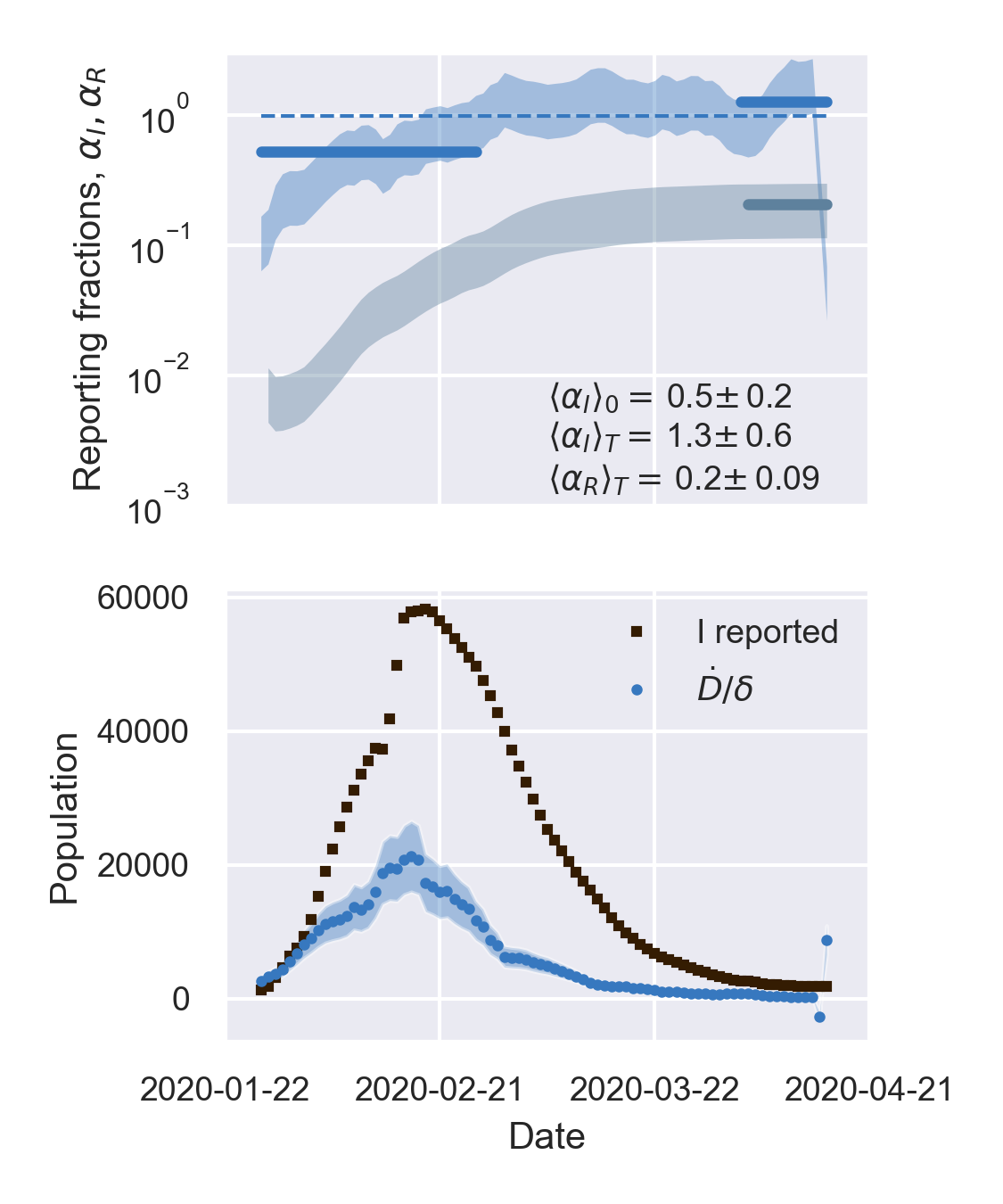}
\end{tabular}
\caption{\textbf{The derivative of the reporting fraction of infected controls the apparent epidemic peak.}
Top panels: fraction of true recovered reported, $\alpha_R(t)$ (grey shaded regions) and 
fraction of true infected reported, $\alpha_I(t)$, computed as described in the text. The shaded
regions correspond to an assumed mortality $m = 0.012 \pm 0.005$~\cite{Bastolla:2020aa}. 
The thick lines mark averages on the last 15 \% portions of the data ($\langle\alpha_{I,R}\rangle_T$), 
and first 40 \% ($\langle\alpha_I\rangle_0$), while the dashed lines
denote the overall average values of $\alpha_{I,R}$ computed over the whole available time span.
Bottom panels. The reported infected are compared to the rescaled deaths/day, 
$I_0(t)~\stackrel{\rm def}{=}~\dot{D}\alpha_I(0)/d$, i.e. the 
infected that would have been reported if the reporting fraction had 
remained constant at its initial value $\alpha_I(0)$. Technically, $\delta = d/\alpha_I(0)$ is 
obtained as the slope $\delta$ of a linear fit of $\dot{D}$ vs $I_M$ in the very early stage of 
the outbreak. 
The shaded regions correspond to the least-square errors found on $\delta$.
As soon as the testing rate starts varying 
appreciably, the two trends $I_0(t)$ and $I_M(t)$ diverge. 
If the testing rate is ramped up, the infected reported 
seem to flag an increase in the outbreak severity (Italy). Conversely, if the testing rate slows 
down (Germany), the reported infected underestimate the true peak. In particular, 
in this case the peak can be alarmingly anticipated. China shows a marginal behaviour, where 
the two peaks coincide, consistently with a remarkably stable reporting fraction of infected.
}
\label{fig:alphadot}
\end{figure}
%
\indent Fig.~\ref{fig:alphadot} illustrates the above considerations for the outbreak 
dynamics in three representative countries, each displaying one of the two characteristic 
behaviors. Technically, in order to compute the reporting fractions $\alpha_I,\alpha_R$, one
needs to know the mortality $m=d/\beta = d/(a+d)$. 
A good country-independent estimate for $m$ has been reported recently by looking 
at the infinite-number of test extrapolation~\cite{Bastolla:2020aa}, 
which yields $m = 0.012$. Hence, combining 
Eqs.~\eqref{eq:SIRD} with the definitions~\eqref{e:alphadef}, it is not difficult to obtain
\begin{equation}
\label{e:alphac}
\alpha_I(t) = m \beta  \frac{I_M(t)}{\dot{D}(t)} \qquad 
\alpha_R(t) = \frac{m}{1-m} \frac{R_M(t)}{D(t)}
\end{equation}
where $\beta$ is estimated from the fits of the SIRD model in the plane $(\dot{D},D)$.
Direct inspection of Fig.~\ref{fig:alphadot} confirms the above reasoning on the 
relative position in time of the apparent and true peaks. For example, it can be seen 
that tests have increased exponentially in Italy, starting from the end of March, which 
caused the apparent peak to occur nearly a month after the true one. Conversely, 
the testing activity has been vigorous in Germany at the onset of the outbreak, 
but appears to have slowed down substantially as the epidemics progressed. 
As a consequence, the apparent peak has been reported while the epidemics was still 
growing.  In both cases, our estimate is consistent with the fact that 
only $15$ \% of infected persons are effectively reported.
China seems to provide an instance of the marginal case $\dot{\alpha}_I = 0$. 
Despite an early ramping up, the reporting fraction of infected appears to be remarkably 
stable, which is consistent with the co-occurence of the true and the apparent peaks. 
According to the data, the testing activity has been resolute since the beginning (only 
about half of the infected eluding tracing), and has culminated 
with a scenario compatible with virtually all infected individuals being traced. \\
\indent The calculations reported in Fig.~\ref{fig:alphadot} also confirm  that counting 
instances of recovery is a much subtler task.  Typically, countries seem to have 
needed some {\em adjustment} time to set up robust counting protocols, as it is 
apparent from the large fluctuations that are often observed in the trends of $\alpha_R$ at early 
times (see also supplementary material). Another interesting feature that emerges clearly from 
the comparison of $\alpha_I$ and $\alpha_R$ is the average recovery time, which can be inferred for 
example by direct inspection of the testing activity in Italy. The curve $\alpha_R(t)$
appears to be shifted in the future by about 10 days with respect to $\alpha_I(t)$, which 
is consistent with the present consensus estimate of an average time for recovery of two weeks.\\
%
\begin{figure}[t!]
\centering
\includegraphics[width=0.7\textwidth]{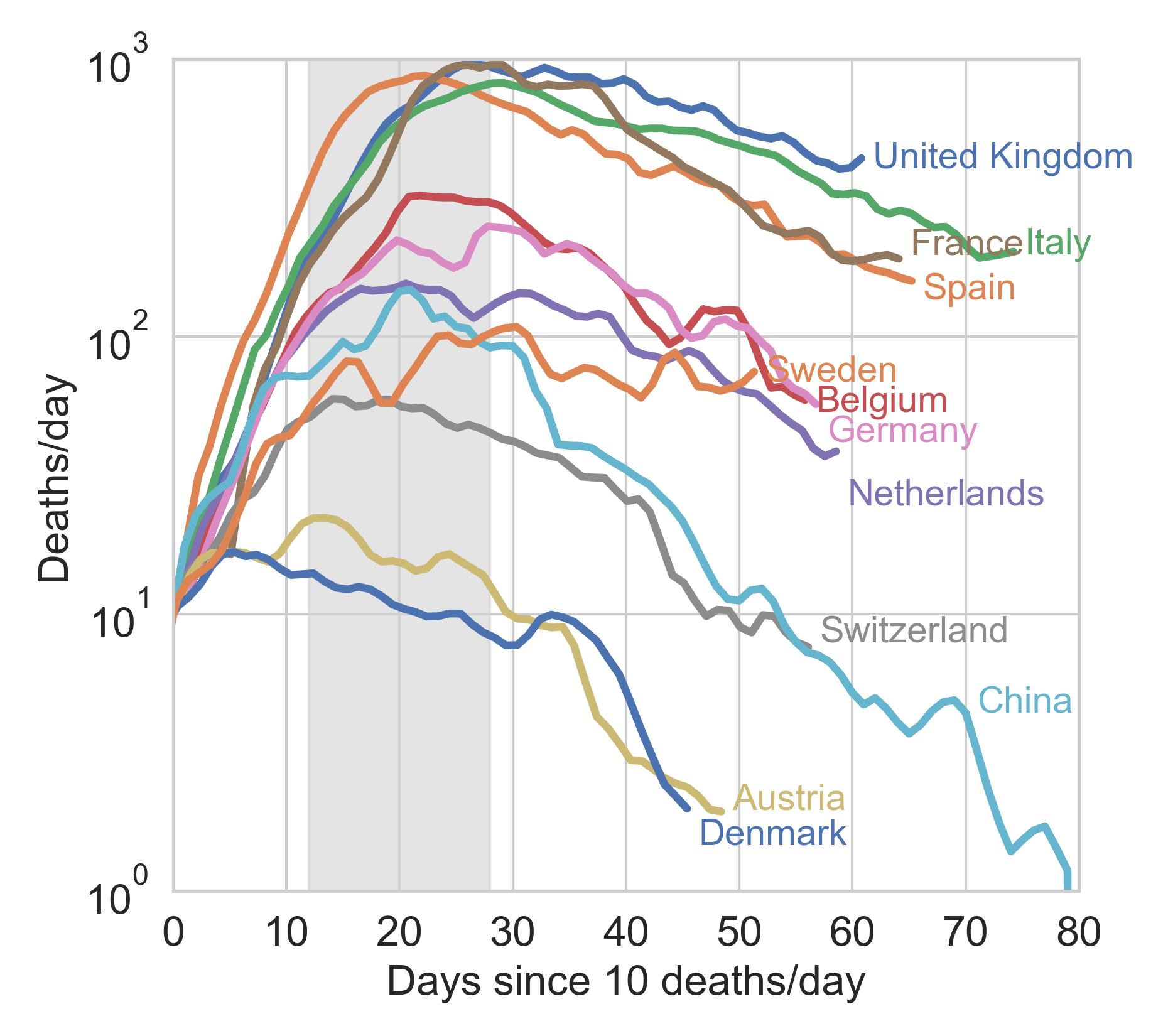}
\caption{\textbf{The true epidemic peak occurs within 30 days after the first ten deaths, the
earlier the higher the initial testing rate.} 
When shifted so that they coincide at the day where ten new deaths were first  reported, the 
$\dot{D}$ time series reveal that the true epidemic peak occurs inevitably rather early
in the outbreak, visibly not later than one month after the first ten casualties reported.
Earlier vigorous testing is obviously key in containing the outbreak (see also discussion in 
the text).}
\label{fig:DDshift}
\end{figure}
%
\indent Our analysis shows how to pinpoint the true peak of the epidemics -- the maximum 
of the deaths/day time series, $\dot{D}$. It is thus natural to investigate how the 
different {\em true} peaks compare among different countries once these time series have 
been brought to a common time origin. This is illustrated in Fig.~\ref{fig:DDshift}
for a number of European countries. It can be clearly appreciated that  the apogee 
of the outbreak in different countries does not occur later than about one month after the 
first ten deaths reported. It is interesting to observe that in countries where the testing rate 
was vigorous at the onset of the outbreak (large $\langle\alpha_I\rangle_0$), 
e.g. Austria, Switzerland, Belgium, 
Germany~\footnote{Netherlands seem to be a remarkable exception to this observation (see
supplementary material).}, the 
peak seems to occur earlier with respect to countries that have started ramping up their 
testing capacity later into the outbreak, such as Italy, Spain, France and U.K. (see 
supporting information). This might just have the obvious explanation that 
early tracking of infected individuals enables more effective isolation from the start, thus leading 
to a less severe evolution of the outbreak as there simply are less contagious people 
able to propagate the disease. It is also interesting to observe that peak position and 
height show some correlation with the initial reproduction numbers measured in the 
$(\dot{D},D)$ plane (see again Fig.~\ref{fig:phiR0x} and Tables S1 and S2
in the supplementary material).
With some notable exceptions (again Netherlands and Denmark), countries that totalised higher 
deaths/day at peak tend to be characterised by higher values of $\mathcal{R}_0$.
Overall, effective and timely testing seems to have been the 
key to successful containment strategies, as it is by now widely advocated~\cite{Peto:2020aa},
as opposed to muscular lockdowns imposed later into the outbreak. 
%

\section{Summary and discussion}

\noindent In summary, the observed behavior of the early 2020 COVID-19 outbreak 
has raised many diverse and puzzling issues in different countries. Many of them have 
adopted drastic containment measures, which had some effect on the spread of the virus but also
left large sectors of the population world-wide 
grappling with the daunting  perspective of pernicious economic recessions. 
On their side,  scientists from all disciplines are having a hard time 
identifying recurring patterns that may help design and tune 
more effective response strategies for the next waves. 
In this letter we have shown that the spread of the epidemics caused 
by this elusive~\cite{Wang:2020aa} pathogen falls into the universality class of SIR models and extensions 
thereof. Based on this fact, an important corollary of our analysis is that
early testing is not only key in making sense of the reported figures, but also 
in controlling the overall severity of the outbreak in terms of casualties. \\
\indent It is intrinsically difficult to disentangle 
the effects imposed by containment measures from the natural habits of a given community 
in terms of social distancing. For example, Italy is recording three times as many deaths per inhabitant
than Switzerland as we write. However, while it is true that Switzerland seemed to have 
tested  nearly ten times as much initially (see supplementary material), 
it is also true that that country has never been put in full lockdown as it was decided in Italy. 
At the same time, Swiss people are known for their law-abiding attitude and social habits that make 
(imposed or suggested) social distancing certainly easier to maintain 
than in southern countries such as Italy. \\
\indent The key message of this letter is that simple models should not be dismissed 
a priori in favour of more complex and supposedly more accurate schemes, which unquestionably 
go into more detail but at the same time rely necessarily on a large number of parameters that 
it is hard to fix unambiguously.  
All in all, we can state with certainty that vigorous early testing and accurate and 
transparent self-evaluation of the testing activity, assuredly virtuous 
attitudes in general, should be considered all the more a priority in the fight against COVID-19.  


\section{Data}

\noindent The data used in this paper span the interval 1/22/20 - 5/16/20 and have been 
retrieved from the github repository associated with the interactive dashboard hosted 
by the Center for Systems Science and Engineering (CSSE) at Johns Hopkins University, 
Baltimore, USA~\cite{Dong:2020aa}.

\section{Methods}
\setcounter{equation}{0}
\renewcommand{\theequation}{M.\arabic{equation}}

\indent In this section we derive Eq.~\eqref{eq:dotDD} and review the definition of the time-dependent 
reproduction rate $\mathcal{R}(t)$. Combining the first and last equation 
in~\eqref{eq:SIRD} we get
\begin{equation}
\label{e:dSdD}
\frac{dS}{dD} = -\frac{r}{d S_0} D
\end{equation}
which can be readily integrated with the initial condition $D(0) = 0$, giving
\begin{equation}
\label{e:St}
S(t) = S_0 e^{-\xi D(t)}
\end{equation}
with $\xi = r/(dS_0)$. Furthermore, since $R(0) = 0$, one has from Eqs.~\eqref{eq:SIRD}
that $R(t) = (a/d) D(t)$.
Taking into account that by definition $I(t) = \dot{D}/d$,  
the conservation law,~$S(t)+I(t)+R(t)+D(t)=S_0+I_0$, can be recast in the following 
form
\begin{equation}
\label{e:SI0}
S_0 + I_0 = S_0 e^{-\xi D} + \frac{\dot{D}}{d} + \left( \frac{a}{d} + 1 \right) D
\end{equation}
Multiplying through by $d$ and taking into account that $S_0 \gg I_0 = \mathcal{O}(1)$, 
yields directly Eq.~\eqref{eq:dotDD} with 
\begin{eqnarray}
\gamma &=& S_0 d             \label{e:gamma} \\ 
\beta  &=& a + d             \label{e:beta} \\
\xi    &=& \frac{r}{dS_0}    \label{e:xi} 
\end{eqnarray}
%

\smallskip

\noindent We now turn to discussing the definition of $\mathcal{R}(t)$. 
The second equation in~\eqref{eq:SIRD} can be rewritten as
\begin{equation}
\dot{I} = \left( a+d\right) \left( \frac{\ds r}{\ds a+d} \frac{\ds S}{\ds S_0}-1\right) I \\
\end{equation}
Following standard convention, we define the time-dependent reproduction number as
\begin{equation}
\label{Rt1}
\mathcal{R}(t)=\frac{\ds r}{\ds a+d} \frac{\ds S}{\ds S_0}
\end{equation}
which, for $t=0$, yields (see Eqs.~\eqref{e:gamma},~\eqref{e:beta},~\eqref{e:xi})
\begin{equation}
\mathcal{R}_0 \stackrel{\rm def}{=} 
\mathcal{R}(0)=\frac{\ds r}{\ds a+d} = \frac{\ds \xi \gamma}{\ds \beta} 
\end{equation}
From~\eqref{eq:SIRD} one readily obtains
\begin{equation}
\label{Rtdef}
\mathcal{R}(t)=1+ \frac{\dot{I}}{\dot{R}+\dot{D}}  
\end{equation}
an expression which can be in principle employed to track the evolution in time of the epidemics. 
Recall, however, that only  $I_M(t)$ and 
$R_M(t)$ are accessible to direct measurements. The associated 
reporting fractions are not known a priori and this may severely bias the analysis if the 
reported figures are used in Eq.~\eqref{Rtdef} or extensions thereof. 
To overcome this limitation, it is expedient to combine the definition~\eqref{Rt1} with Eq.~\eqref{e:St}
to  obtain a closed expression for  
$\mathcal{R}$ that only depends on $D$, namely
\begin{equation}
\label{Rt}
\mathcal{R}(D)= \mathcal{R}_0 \, e^{-\xi D} 
\end{equation}
Eq.~\eqref{Rt} can be used to estimate the time evolution of the reproduction number
without the bias introduced by the non-stationarity of the reporting activity. 	
In the supplementary material, it is shown that the curve of the infected $I$ 
peaks at $\overline{D}=\gamma \log \mathcal{R}_0 / \beta \mathcal{R}_0$. Plugging
the latter expression into the right-hand side of Eq.~\eqref{Rt} yields
$\mathcal{R}(\overline{D})=1$,  as it should for consistency requirements. 
%

\section{Supplementary information}

\setcounter{equation}{0}
\renewcommand{\theequation}{S.\arabic{equation}}

%
\begin{table}[h!]
\centering
\begin{small}
\caption{\small Best-fit values of the fitting parameters in the $(\dot{D},D)$ plane $\gamma,\beta,\xi$ and 
of the linear fit $\dot{D} = \delta I_M$. 
The errors reported have been computed by constructing confidence ellipses at 90 \% confidence level ($\gamma,\beta,\xi$),
and as standard least-squares errors for the linear fit to extract the slope $\delta$.}
\begin{tabular}[t]{lrrrr}
\toprule
{\bf Country}    &  $\gamma \pm \Delta\gamma$ (days$^{-1}$)
                                         & $\beta \pm \Delta\beta$ (days$^{-1}$)     
                                                                        & $\xi \pm \Delta\xi$   & $\delta \pm \Delta\delta$ (days$^{-1}$).  \\
\midrule
Belgium          &    737 $ \pm $  6     &    0.0751 $ \pm $ 0.0006     &   0.000426  $ \pm $ 4$\times 10^{-6}$  &   0.013 $ \pm $ 0.002    \\
Portugal         &     56 $ \pm $  3     &    0.038  $ \pm $ 0.003      &   0.0046    $ \pm $ 0.0003             &   0.004 $ \pm $ 0.001    \\
Austria          &     51 $ \pm $  6     &    0.075  $ \pm $ 0.009      &   0.0059    $ \pm $ 0.0008             &   0.0012$ \pm $ 0.0003   \\
Switzerland      &    136 $ \pm $  5     &    0.067  $ \pm $ 0.002      &   0.0022    $ \pm $ 9$\times 10^{-5}$  &   0.0026$ \pm $ 0.0005   \\
Sweden           &    125 $ \pm $  2     &    0.0177 $ \pm $ 0.0008     &   0.0018    $ \pm $ 5$\times 10^{-5}$  &   0.009 $ \pm $ 0.002    \\
Italy            &   1245 $ \pm $  2     &    0.0333 $ \pm $ 0.0001     &   0.0002206 $ \pm $ 5$\times 10^{-7}$  &   0.016 $ \pm $ 0.004    \\
France           &   2086 $ \pm $  5     &    0.0713 $ \pm $ 0.0001     &   0.0001572 $ \pm $ 4$\times 10^{-7}$  &   0.006 $ \pm $ 0.001    \\
Spain            &   1312 $ \pm $  2     &    0.0426 $ \pm $ 0.0001     &   0.0002946 $ \pm $ 6$\times 10^{-7}$  &   0.016 $ \pm $ 0.004    \\
Germany          &    500 $ \pm $  5     &    0.0537 $ \pm $ 0.0006     &   0.000518  $ \pm $ 6$\times 10^{-6}$  &   0.001 $ \pm $ 0.0002   \\
Denmark          &     24 $ \pm $  2     &    0.036  $ \pm $ 0.004      &   0.013     $ \pm $ 0.002              &   0.005 $ \pm $ 0.001    \\
U. K.            &   1483 $ \pm $  2     &    0.0323 $ \pm $ 0.0001     &   0.0001919 $ \pm $ 4$\times 10^{-7}$  &   0.019 $ \pm $ 0.005    \\
Netherlands      &    246 $ \pm $  2     &    0.035  $ \pm $ 0.0004     &   0.00121   $ \pm $ 2$\times 10^{-5}$  &   0.012 $ \pm $ 0.003    \\
Turkey           &    283 $ \pm $  8     &    0.058  $ \pm $ 0.002      &   0.00086   $ \pm $ 2$\times 10^{-5}$  &   0.0041$ \pm $ 0.0009   \\
Iran             &    209 $ \pm $  1     &    0.0229 $ \pm $ 0.0002     &   0.00128   $ \pm $ 1$\times 10^{-5}$  &   0.011 $ \pm $ 0.003    \\
US               &   2662 $ \pm $  2     &    0.0135 $ \pm $ 0.0001     &   0.0001021 $ \pm $ 1$\times 10^{-7}$  &   0.006 $ \pm $ 0.001    \\
China            &    670 $ \pm $ 30     &    0.168  $ \pm $ 0.005      &   0.00055   $ \pm $ 2$\times 10^{-5}$  &   0.006 $ \pm $ 0.002    \\
\bottomrule
\end{tabular}
\end{small}
\end{table}%

%
\begin{table}[h!]
\centering
\begin{small}
\caption{\small Number of deaths at the true infection peak, $\overline{D}$, computed 
self-consistently as $\overline{D}=\gamma \log \mathcal{R}_0 / \beta \mathcal{R}_0$ 
(see Eq.~\eqref{eq:Dhat}) and estimated values for the {\em effective} basin of 
susceptible $S_0$, computed as $S_0 = \gamma/d = \gamma/(\delta \langle \alpha_I\rangle_0)$. 
The averages in $\langle \alpha_I\rangle_0$ correspond to the first 40 \% of the data 
(starting from 22/01/2020), see also Fig. 2 in the main text.
A plot of $\overline{D}$ vs $S_0$ is reported in Fig.~\ref{fig:DmaxvsS0}.
\label{t:Dmax}}
\begin{tabular}[t]{lrr}
\toprule
{\bf Country}    &  $\overline{D} \pm \Delta\overline{D}$ 
                                        & $S_0\pm \Delta S_0$             \\
\midrule
Belgium          &   3360 $ \pm $  80	&     200000 $ \pm $  100000      \\
Portugal         &    400 $ \pm $ 100	&      70000 $ \pm $   60000	  \\
Austria          &    240 $ \pm $  80	&      50000 $ \pm $   50000	  \\
Switzerland      &    680 $ \pm $  70	&     200000 $ \pm $  100000	  \\
Sweden           &   1400 $ \pm $ 200	&      80000 $ \pm $   60000	  \\
Italy            &   9560 $ \pm $  60	&    3000000 $ \pm $ 2000000	  \\
France           &   9710 $ \pm $  70	&    3000000 $ \pm $ 2000000	  \\
Spain            &   7480 $ \pm $  40	&    1000000 $ \pm $ 1000000	  \\
Germany          &   3000 $ \pm $ 100	&     600000 $ \pm $  400000	  \\
Denmark          &    170 $ \pm $  60	&      10000 $ \pm $   10000	  \\
U. K.            &  11340 $ \pm $  90	&    1300000 $ \pm $  800000	  \\
Netherlands      &   1770 $ \pm $  70	&     300000 $ \pm $  200000	  \\
Turkey           &   1700 $ \pm $ 100	&     500000 $ \pm $  400000	  \\
Iran             &   1920 $ \pm $  70	&    1200000 $ \pm $  800000	  \\
US               &  29300 $ \pm $ 100   &   15000000 $ \pm $ 9000000	  \\
China            &   1400 $ \pm $ 100	&     200000 $ \pm $  100000	  \\
\bottomrule
\end{tabular}
\end{small}
\end{table}%

%
\begin{figure}[h!]
\centering
\includegraphics[width=0.6\textwidth]{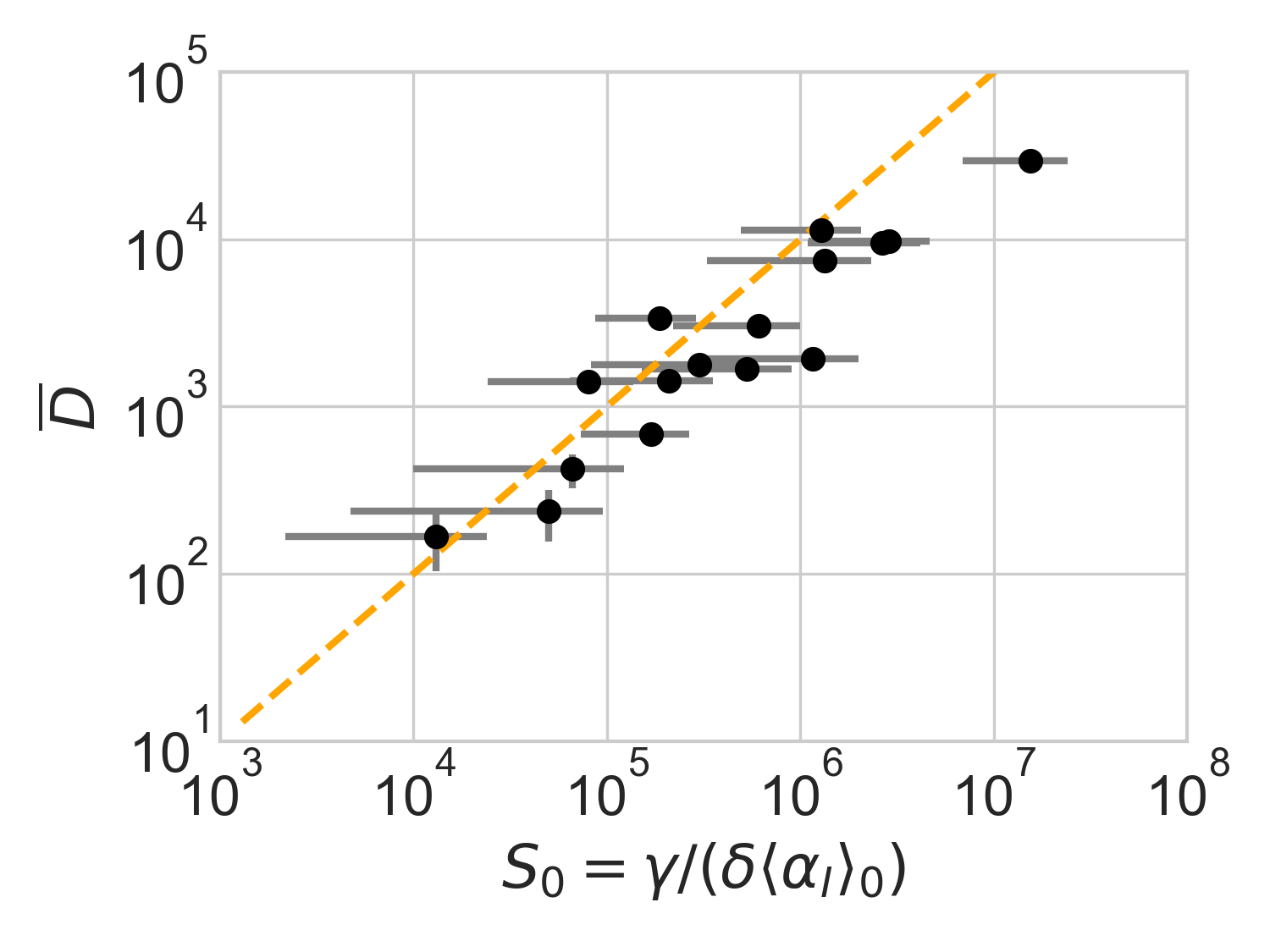}
\caption{\textbf{Number of deaths at the true infection peak vs effective population size}
estimated self-consistently (see Table~\ref{t:Dmax}). 
A direct proportionality is observed, in agreement with relation~\eqref{eq:Dhat}. 
The vertical and horizontal bars denote  the errors reported in Table~\ref{t:Dmax}, 
calculated according to standard error propagation from the formulae used to compute the two 
quantities. The orange dashed line is the arbitrary proportionality relation $\overline{D}= S_0/100$,
drawn to provide an order of magnitude of the ratio between deaths at peak and effective 
size of the infection basin.}
\label{fig:DmaxvsS0}
\end{figure}
%

\clearpage

\subsection{Infection peak}

\indent In this section we derive some useful relations about key quantities at the 
infection peak, namely the time $\overline{t}$ at which the number of infected persons, 
$I(t)$, reaches its maximum value, $\overline{I}=I(\overline{t})$.
This can be determined from the condition $\dot{I}(\overline{t})=0$. 
From the fourth equation of~(1) in the main text, we get $\ddot{D}(\overline{t})=0$, 
then using the definition of $f(D)$ we can write
\begin{equation*}
0=\ddot{D}(\overline{t}) = f'(\overline{D})\dot{D}(\overline{t})\, ,
\end{equation*}
where $\overline{D}=D(\overline{t})$.
Being $D$ a monotonic increasing function we have, $\dot{D}(t)>0$ for all $t$, 
hence the latter condition is equivalent to
\begin{equation*}
0= f'(D(\overline{t}))\, ,
\end{equation*}
where the $'$ stand for the derivative of $f$ with respect to the variable $D$. 
This equation can be straightforwardly solved to give
\begin{equation}
\label{eq:Dhat}
\overline{D} = \frac{dS_0}{r} \log \frac{r}{a+d}=S_0\frac{d}{a+d} \frac{1}{\mathcal{R}_0}\log \mathcal{R}_0 \, ,
\end{equation}
where  $\mathcal{R}_0=r/(a+d)$ (see also Methods in the main text).
We are now able to compute the remaining key quantities at the peak time $\overline{t}$. 
Indeed, the relations developed in  the Methods section in the main text imply
\begin{eqnarray*}
\overline{R}&=&S_0\frac{a}{a+d} \frac{1}{\mathcal{R}_0}\log \mathcal{R}_0\\
 \overline{S}&=& S_0\frac{1}{\mathcal{R}_0}\\
 \overline{I} &=& S_0\left(1-\frac{1}{\mathcal{R}_0}- \frac{1}{\mathcal{R}_0}\log \mathcal{R}_0\right)\, .
\end{eqnarray*}
Let us observe that these quantities, as well as $\overline{D}$, scale proportionally to 
the system size, $S_0$, as one can appreciate from Fig.~\ref{fig:DmaxvsS0} where we report 
$\overline{D}$ as a function of $S_0$. The latter is computed as $S_0 = \gamma/d$
(see main text). Here $d = \delta \alpha_I(0)$, where $\delta$ is the slope of a linear fit  
$\dot{D}$ vs $I_M$ at short times and $\gamma$ is a parameter of the function $f(D)$ 
obtained from a fit in the plane $(\dot{D},D)$ (see Fig. 1 in the main text).\\
%

\subsection{Rescaled variables}

\begin{figure}[t!]
\centering
\includegraphics [width=0.85\textwidth]{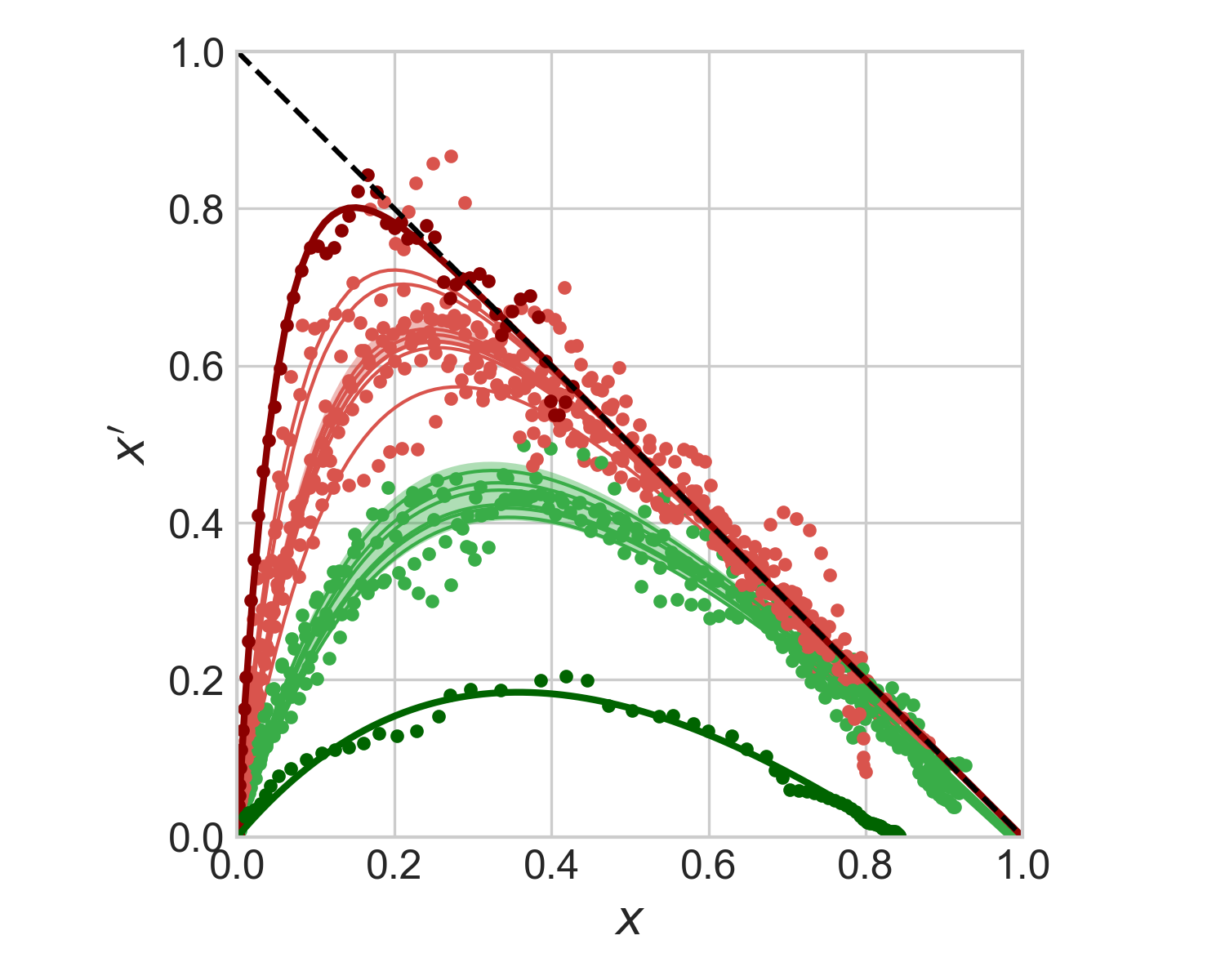}
\caption{\textbf{Universal behaviour of the COVID-19 outbreak in the rescaled plane
$(x^\prime,x)$}. 
Symbols stand for the data $\dot{D},D$ 
rescaled according to the prescriptions~\eqref{eq:rescvar1} and~\eqref{eq:rescvar},
respectively. The green and red shaded regions highlight the intervals $4\leq\mathcal{R}_0\leq5$ 
and $8.5\leq\mathcal{R}_0\le9.5$, respectively. The thick green and red lines denote
the data from the outbreaks in China ($\mathcal{R}_0 = 2.2 \pm 0.2$) 
and the United States  ($\mathcal{R}_0 = 20.0 \pm 0.1$), respectively.
}
\label{fig:dotxx}
\end{figure}

\noindent In the main text we have shown (see Fig.1) that the epidemic spreading of the COVID-19 
falls for a large number of countries in the SIR universality class, 
once read in the plane $(\dot{D},D)$. Indeed, all 
the data fit very well on the curve $\dot{D}=f(D)$, the latter depending on three parameters.
One can go one step forward and show that there is a particular choice of rescaled variables such 
that all  the curves $\dot{D}=f(D)$ collapse on a one-parameter family of curves, 
indexed by the effective basic reproduction number, $\mathcal{R}_0$.
Let the define a new variable $x$ and a new time $t'$ by
\begin{equation}
\label{eq:rescvar}
x = \frac{a+d}{d}\frac{D}{N} = \frac{\beta D}{\gamma}   \qquad t' = (a+d) t = \beta t\, ,
\end{equation}
where $N = S_0 + I_0 \approx S_0$. Then (see Eq.~(2) in the main text and Methods)
\begin{eqnarray}
\label{eq:rescvar1}
x^\prime&=&\frac{dx}{dt'}=\frac{dx}{dt}\frac{dt}{dt'}=
\frac{\dot{D}}{\gamma} \\
 &=& 1- e^{-\xi D(t)}  -\frac{\beta D(t)}{\gamma}\notag \nonumber\\
 &=&1-e^{-\mathcal{R}_0x}-x \stackrel{\rm def}{=} \phi(x;\mathcal{R}_0)
\end{eqnarray}
where we have used the relation $\mathcal{R}_0 = \xi\gamma/\beta$.
The universal function is unimodal, $\phi(0;\mathcal{R}_0)=0$ for all values of 
$\mathcal{R}_0$ and there is a single positive root, 
$\tilde{x}(\mathcal{R}_0)$, which converges to $1$ for $\mathcal{R}_0\rightarrow \infty$. 
The function $\phi$ is linear for small values of $x$ with slope  
$\mathcal{R}_0-1$, while at $\tilde{x}(\mathcal{R}_0)$ one  has
\[
\phi'(\tilde{x}(\mathcal{R}_0);\mathcal{R}_0) = \mathcal{R}_0 [1-\tilde{x}(\mathcal{R}_0)]-1
\]
It is easy to see that $\phi'(\tilde{x}_{\mathcal{R}_0};\mathcal{R}_0)\rightarrow -1$ as 
$\mathcal{R}_0\rightarrow \infty$. \\
\indent Fig.~\ref{fig:dotxx} reports the same data as plotted in Fig.~1 in the main text, but
in the rescaled plane $(x^\prime,x)$. It is apparent that there exist two main clusters 
of outbreaks, one with values of $\mathcal{R}_0$ between 4 and 5 and one 
with values of $\mathcal{R}_0$ between 8.5 and 9.5. The outbreak in China is the 
only observed case where $\tilde{x}(\mathcal{R}_0)<1$, due to the low value 
of $\mathcal{R}_0 = 2.2 \pm 0.2$ computed from the fit.   

\subsection{Non-stationary reduction of the infection rate and the effective reproduction number}

\begin{figure}[t!]
\centering
\includegraphics [width=\textwidth]{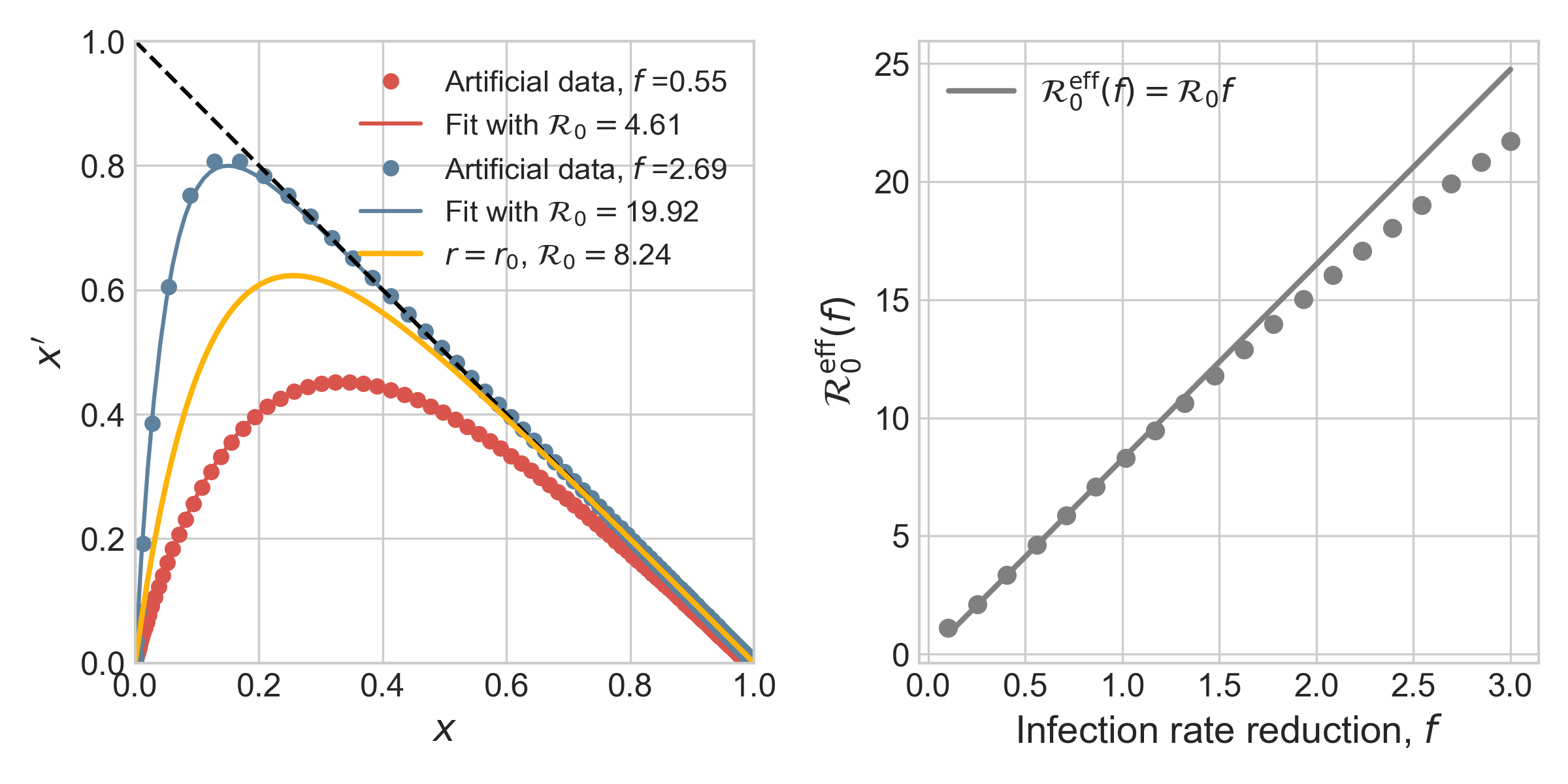}
\caption{\textbf{SIRD-like behaviour in the rescaled plane
$(x^\prime,x)$ is observed for non-stationary infection rate with an 
effective reproduction number}. 
Starting from a given set of parameters, 
$r_0=0.275$ days$^{-1}$, $a = 0.0309$ days$^{-1}$, $d = 0.00235$ days$^{-1}$,
corresponding to $\mathcal{R}_0 = 8.24$,
we have generated different {\em artificial} SIRD dynamics starting from the same initial conditions,
$S_0 = 528747$, $I_0=1$ according to Eq.~\eqref{e:rt} with $f \in [0.1,3]$, $t^\ast = 20$ days
and $\Delta t = 3.5$ days. 
Each non-stationary SIRD dyamics has been fitted in the $(x^\prime,x)$  plane
with the function $\phi(x;\mathcal{R}^{\rm eff}_0) = 1 - \exp(-\mathcal{R}^{\rm eff}_0 x) -x$
(two instances are plotted in the left panel). The effective reproduction numbers 
are plotted in the right panel against the corresponding reduction of infection rate, $f$.
These seem to be described by the same universal linear reduction, 
$\mathcal{R}^{\rm eff}_0(f) = \mathcal{R}_0 f$, irrespective of the 
characteristic time scales that govern the non-stationarity and even of the 
specific function the drives the reduction from $r_0$ to $f r_0 < r_0$ (see text). 
}
\label{fig:R0eff}
\end{figure}

\noindent The reproduction numbers measured from the fits performed in the $(\dot{D},D)$ plane 
should be considered as {\em effective} values. These most likely incorporate all 
sources of non-stationarity left besides the time-dependent bias associated with the 
evolution of the testing rate. In particular, when imposed on the population, 
the effect of a lockdown is that of reducing over a characteristic (possibly rather short)
period  of time the rate of infection $r$. The question then arises whether the portrait 
in the $(\dot{D},D)$ plane of a SIRD evolution where 
the rate of infection $r$ is a time-dependent function will still be described
by the same universal function $f(D)$ but with an {\em effective} reproduction 
number $\mathcal{R}_0^{\rm eff}$. It turns out that the answer is in the affirmative 
and the rationale for this is deceptively simple.\\
\indent In order to explore this question, we consider a modified 
version of the SIRD model (Eqs. (1) in the main text), where the  infection rate $r$ 
is let vary with time~\cite{Fanelli:2020aa}. 
More precisely, if some containment measures were enforced at time $t^\ast$,
causing their effect on a typical time $\Delta t$, we may take
\begin{equation}
\label{e:rt}
r(t) = \begin{cases}
       r_0 & \text{for} \quad t\leq t^\ast \\
       r_0(1-f) e^{-(t-t^\ast)/\Delta t} + f \, r_0 &
       \text{for} \quad t >  t^\ast
      \end{cases} 
\end{equation}
where $r_0$ is the initial, pre-lockdown infection rate that defines 
the {\em true} initial reproduction number, $\mathcal{R}_0 = r_0/(a+d)$. 
The fraction $f$ gauges the asymptotic reduction (if $f<1$) of the infection rate 
afforded by the containment measures. Of course, this kind of description can 
be easily adapted to the case where the non-stationarity causes the opposite effect, 
that is, $f>1$. This might be the case, for example, of overlooked hotbeds or 
large undetected gatherings that cause the infection to accelerate on average 
at a certain point in time. \\ 
\indent Whatever the effect of the non-stationarity introduced by $r(t)$, the 
effect in the rescaled plane $(x^\prime,x)$ turns out to be extremely simple and 
apparently  robust with respect to 
the choice of the characteristic time scales $t^\ast,\Delta t$
and also of the specific functional form that describes the reduction from 
$r_0$ to $f r_0$. This is illustrated in Fig.~\ref{fig:R0eff} for a choice of 
$f$ in the interval $[0.1,3]$.
%
%
The evolution of outbreaks generated from a given {\em root} one, that is, one with a given 
infection rate $r(0) = r_0$, can always be described by the same
one-parameter function  $\phi(x;\mathcal{R}^{\rm eff}_0)$ with an effective reproduction
number $\mathcal{R}^{\rm eff}_0$ that only depends on $f$. 
Moreover, $\mathcal{R}^{\rm eff}_0$ for $f<1$ is 
simply the $t=0$ value reduced by the same  factor, that is 
\begin{equation}
\label{e:R0eff}
\mathcal{R}^{\rm eff}_0(f) = \mathcal{R}_0 f
\end{equation}
Remarkably, this conclusion, for $f<1$, does not depend on the particular choice of 
the time scale $\Delta t$, nor on the onset time $t^\ast$, nor 
on the particular choice for the function that describes the 
fading of the infection rate. For example, a function  of the kind $1/(1+x)$ instead 
of an exponential damping in Eq.~\eqref{e:rt} yields exactly the same result~\eqref{e:R0eff},
that we might then consider a general fact.

\newpage

\begin{figure}[t!]
\centering
\begin{tabular}{ccc}
Belgium & Portugal & Spain\\
\includegraphics [width=0.33\textwidth]{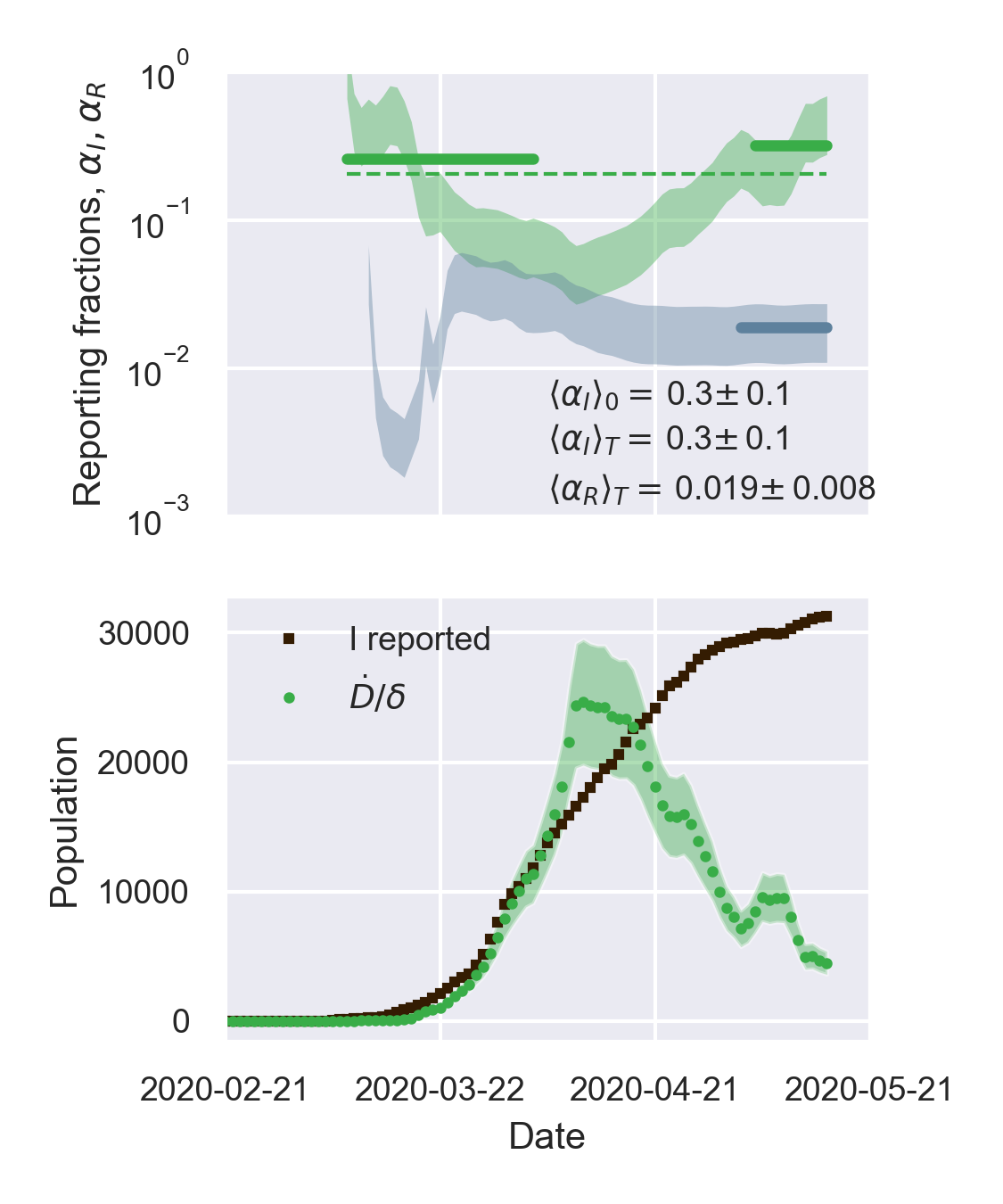} &  
\includegraphics [width=0.33\textwidth]{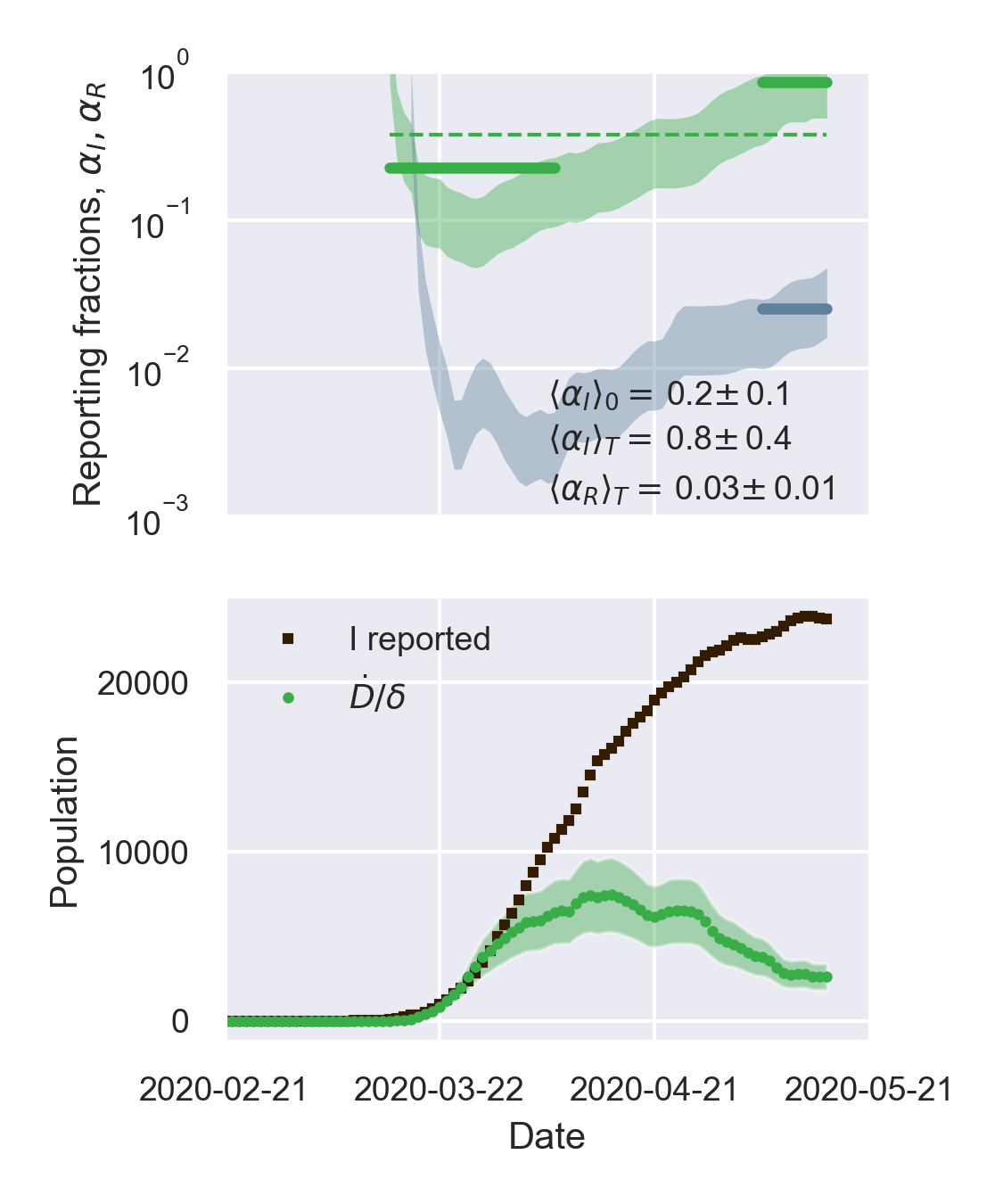} & 
\includegraphics [width=0.33\textwidth]{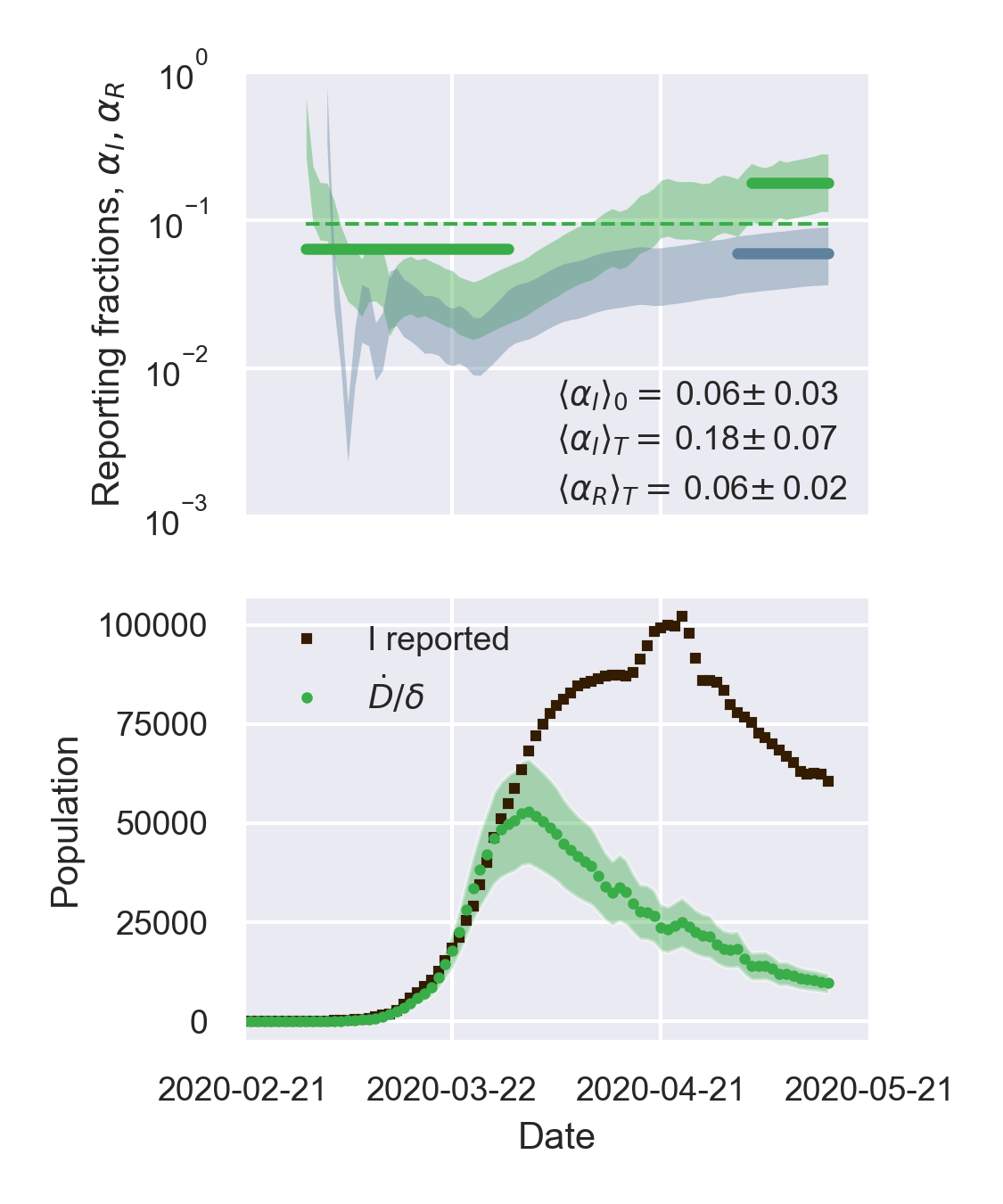} \\
Netherlands & United Kingdom & United States\\
\includegraphics [width=0.33\textwidth]{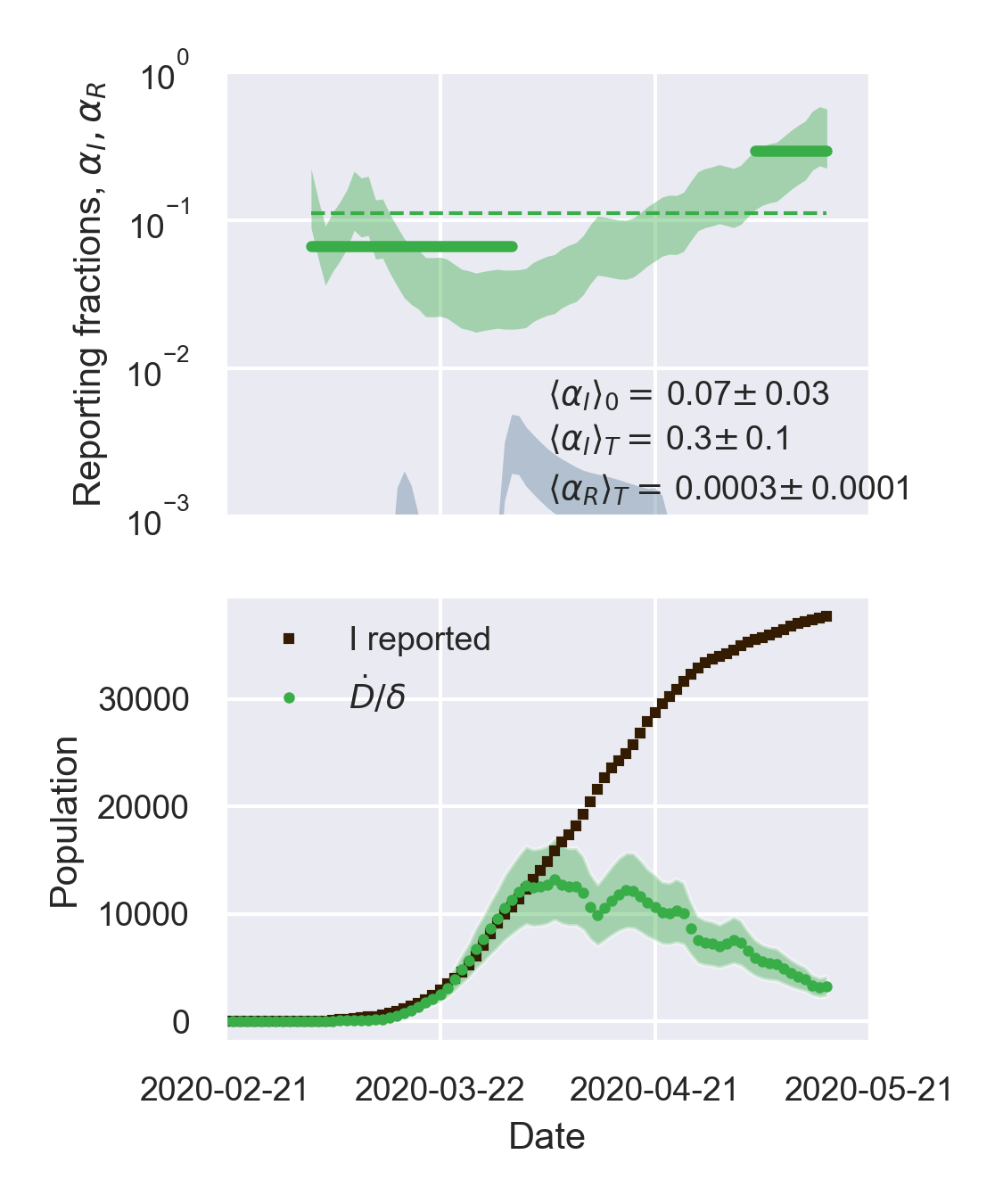} &  
\includegraphics [width=0.33\textwidth]{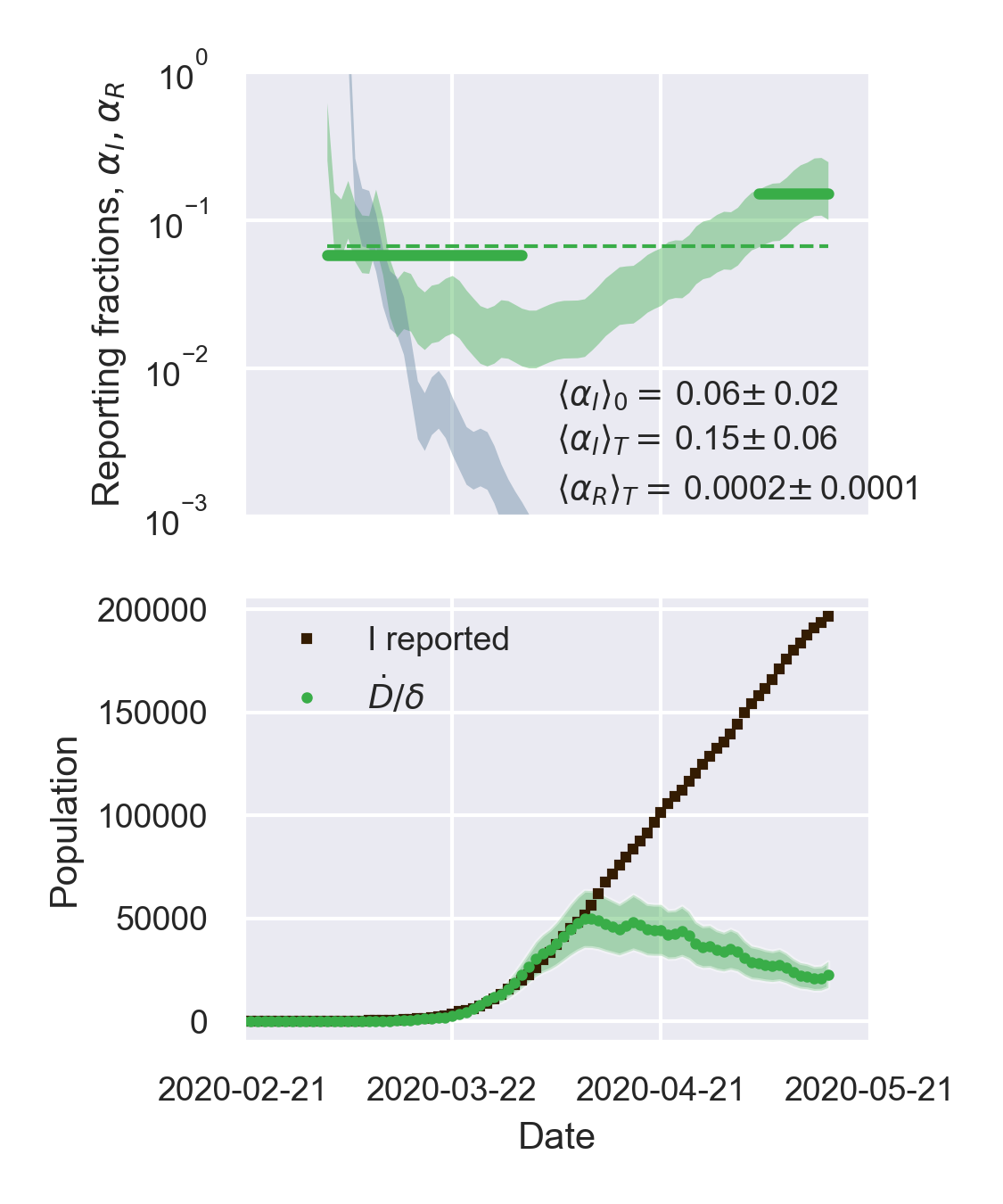} & 
\includegraphics [width=0.33\textwidth]{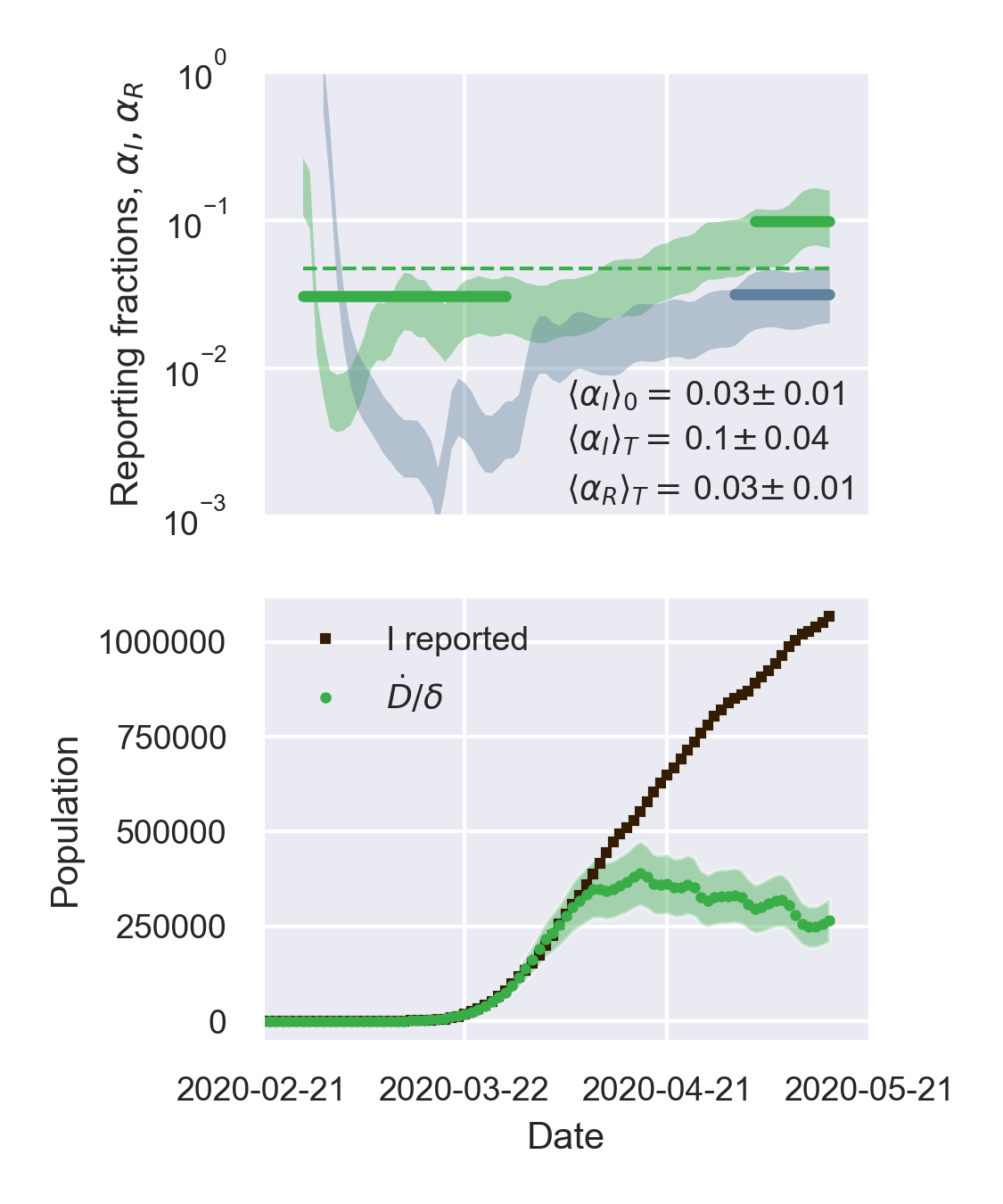} 
\end{tabular}
\caption{\textbf{A positive derivative of the reporting fraction of infected causes a delay of
the apparent epidemic peak with respect to the true one - I.}
Top panels: fraction of true recovered reported, $\alpha_R(t)$ (grey shaded regions) and 
fraction of true infected reported, $\alpha_I(t)$, computed as described in the text. The shaded
regions correspond to an assumed mortality $m = 0.012 \pm 0.005$~\cite{Bastolla:2020aa}. 
The thick lines mark averages on the last 15 \% portions of the data ($\langle\alpha_{I,R}\rangle_T$), 
and first 40 \% ($\langle\alpha_I\rangle_0$), while the dashed lines
denote the overall average values of $\alpha_{I,R}$ computed over the whole available time span.
Bottom panels. The reported infected are compared to the rescaled deaths/day, 
$I_0(t)~\stackrel{\rm def}{=}~\dot{D}\alpha_I(0)/d$, i.e. the 
infected that would have been reported if the reporting fraction had 
remained constant at its initial value $\alpha_I(0)$. Technically, $\delta = d/\alpha_I(0)$ is 
obtained as the slope $\delta$ of a linear fit of $\dot{D}$ vs $I_M$ in the very early stage of the outbreak.
The shaded regions corresponds to the least-square errors found on $\delta$.
}
\label{fig:alphadot11}
\end{figure}
 
\begin{figure}[t!]
\centering
\begin{tabular}{ccc}
Sweden & Turkey & Iran\\
\includegraphics [width=0.33\textwidth]{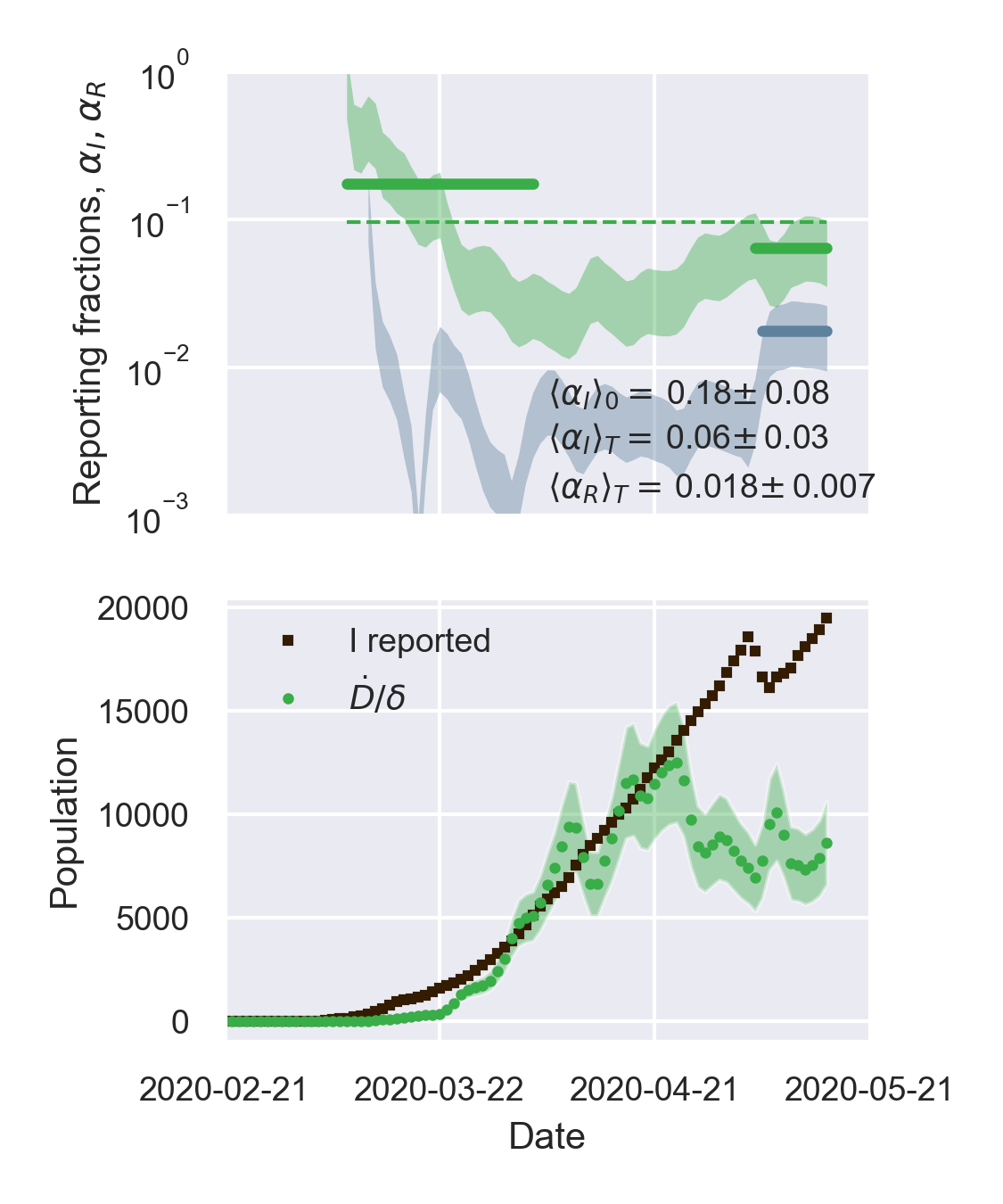} &  
\includegraphics [width=0.33\textwidth]{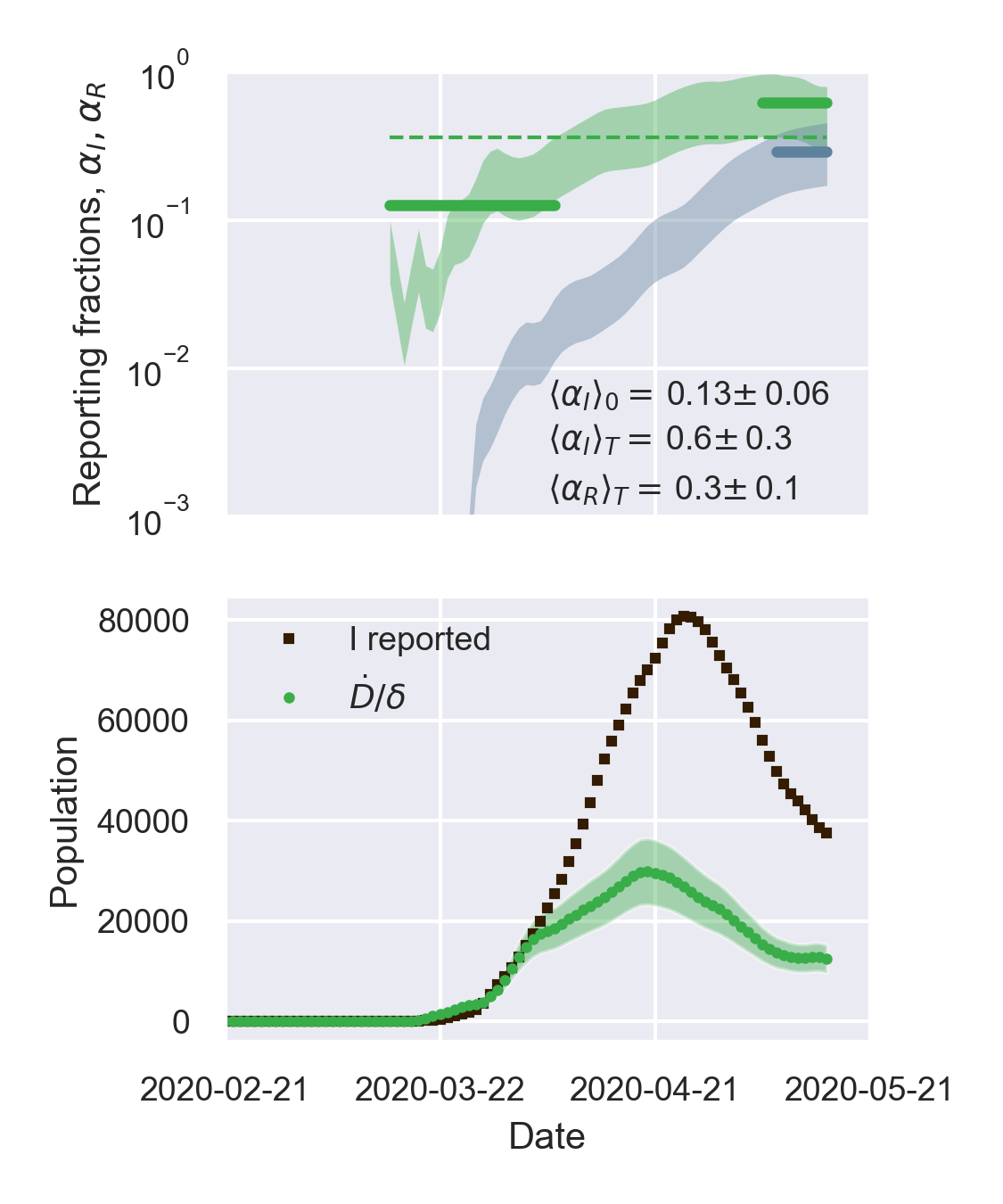} & 
\includegraphics [width=0.33\textwidth]{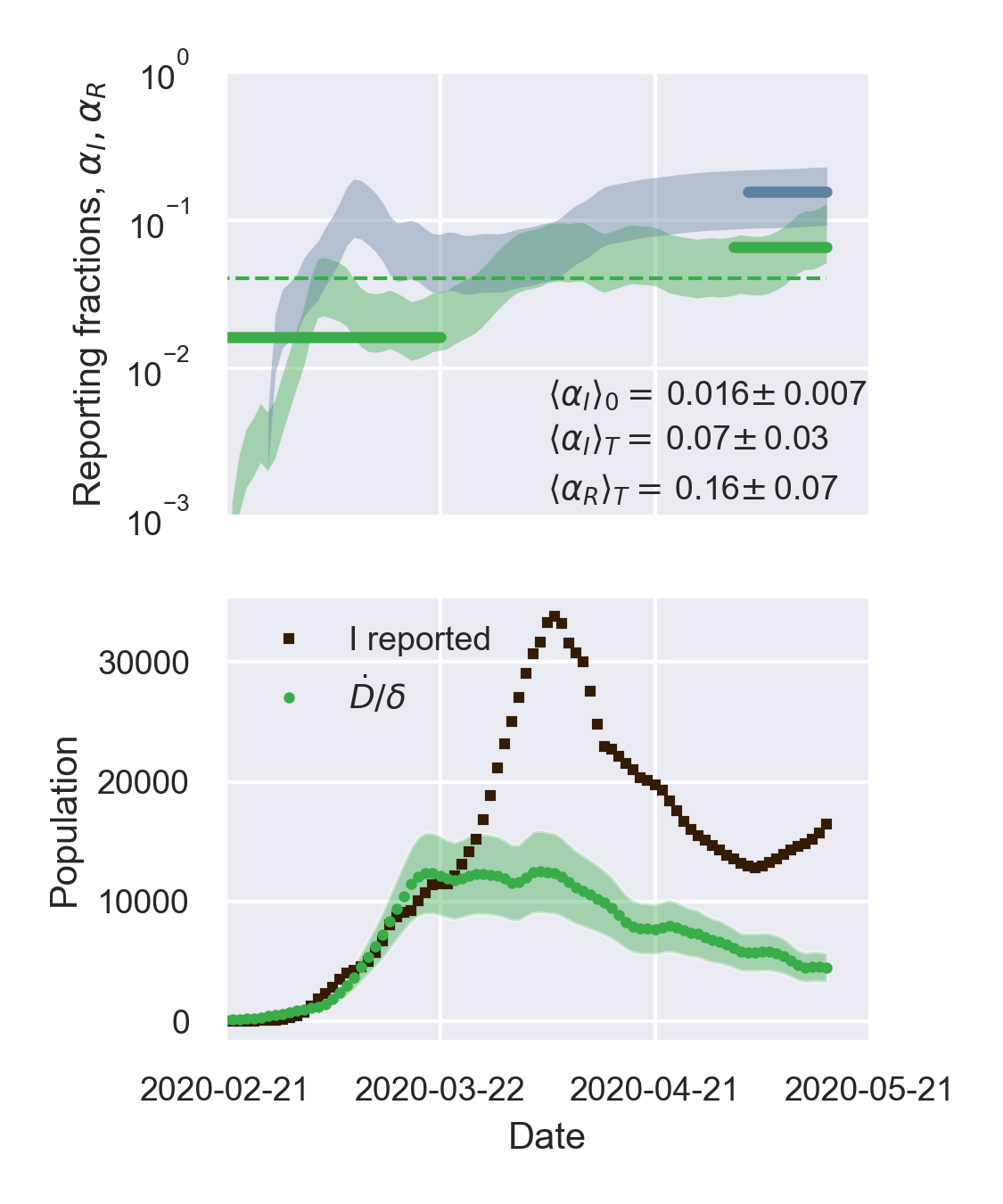} \\
Greece &  Indonesia & Philippines \\
\includegraphics [width=0.33\textwidth]{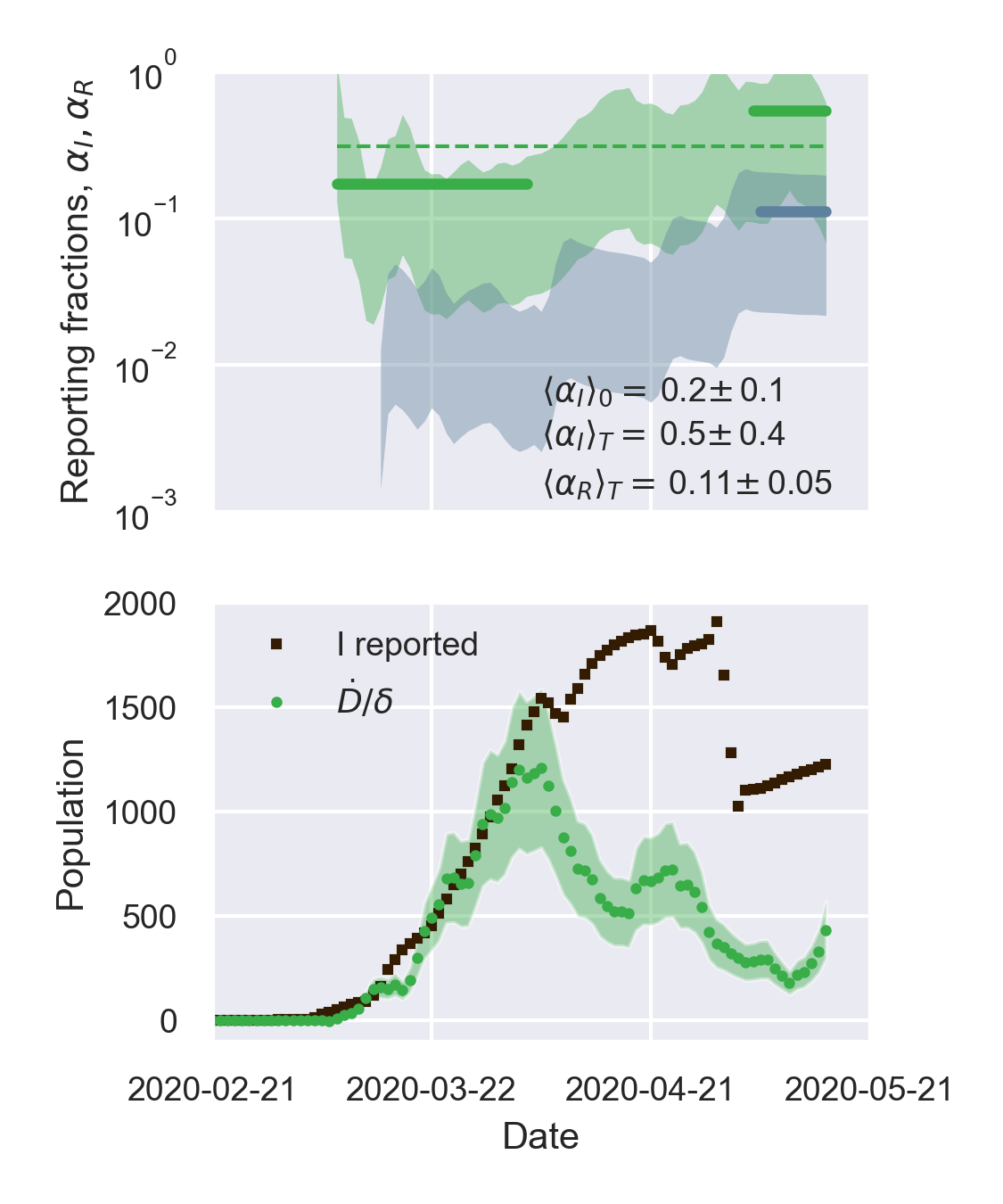} &  
\includegraphics [width=0.33\textwidth]{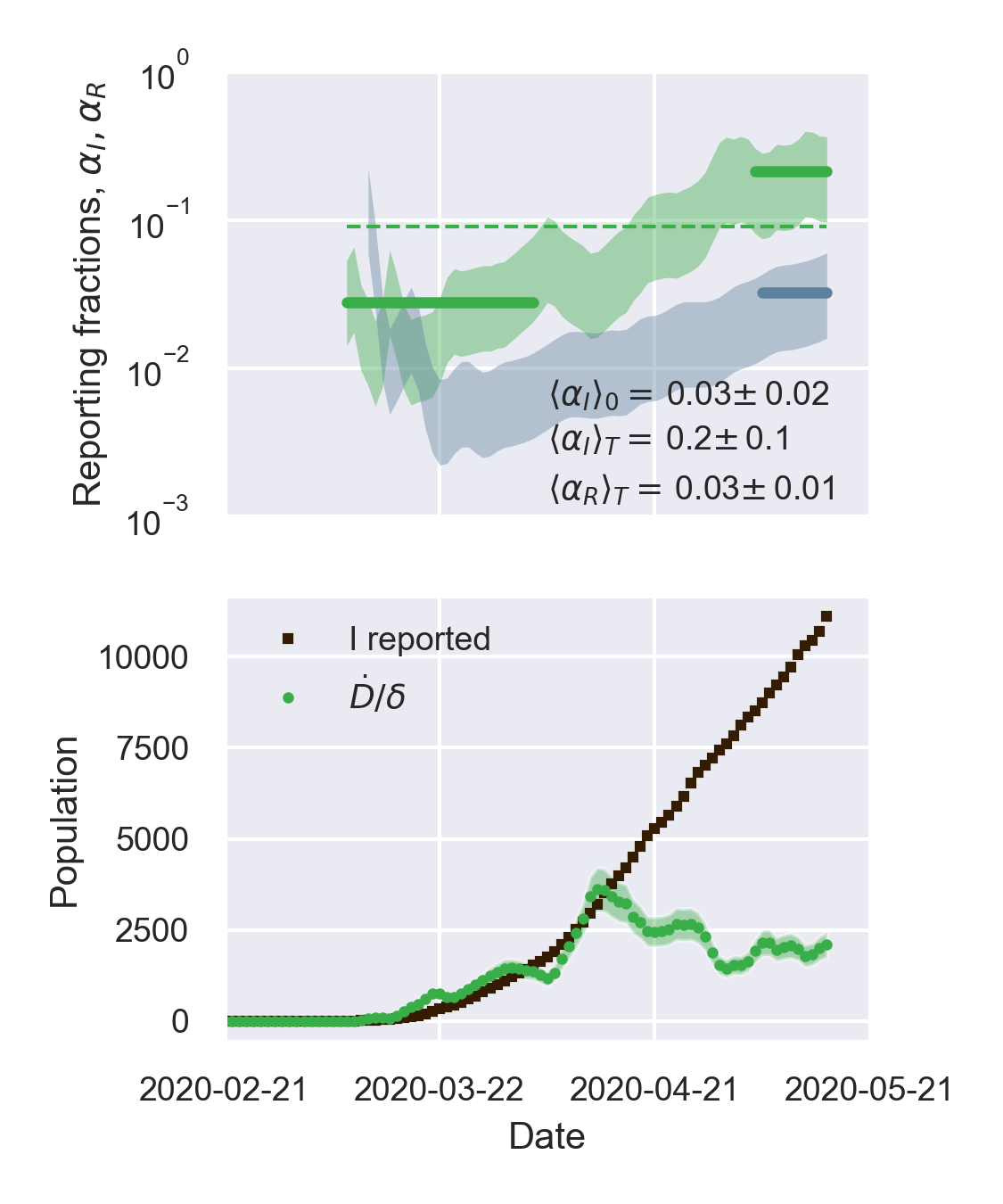} & 
\includegraphics [width=0.33\textwidth]{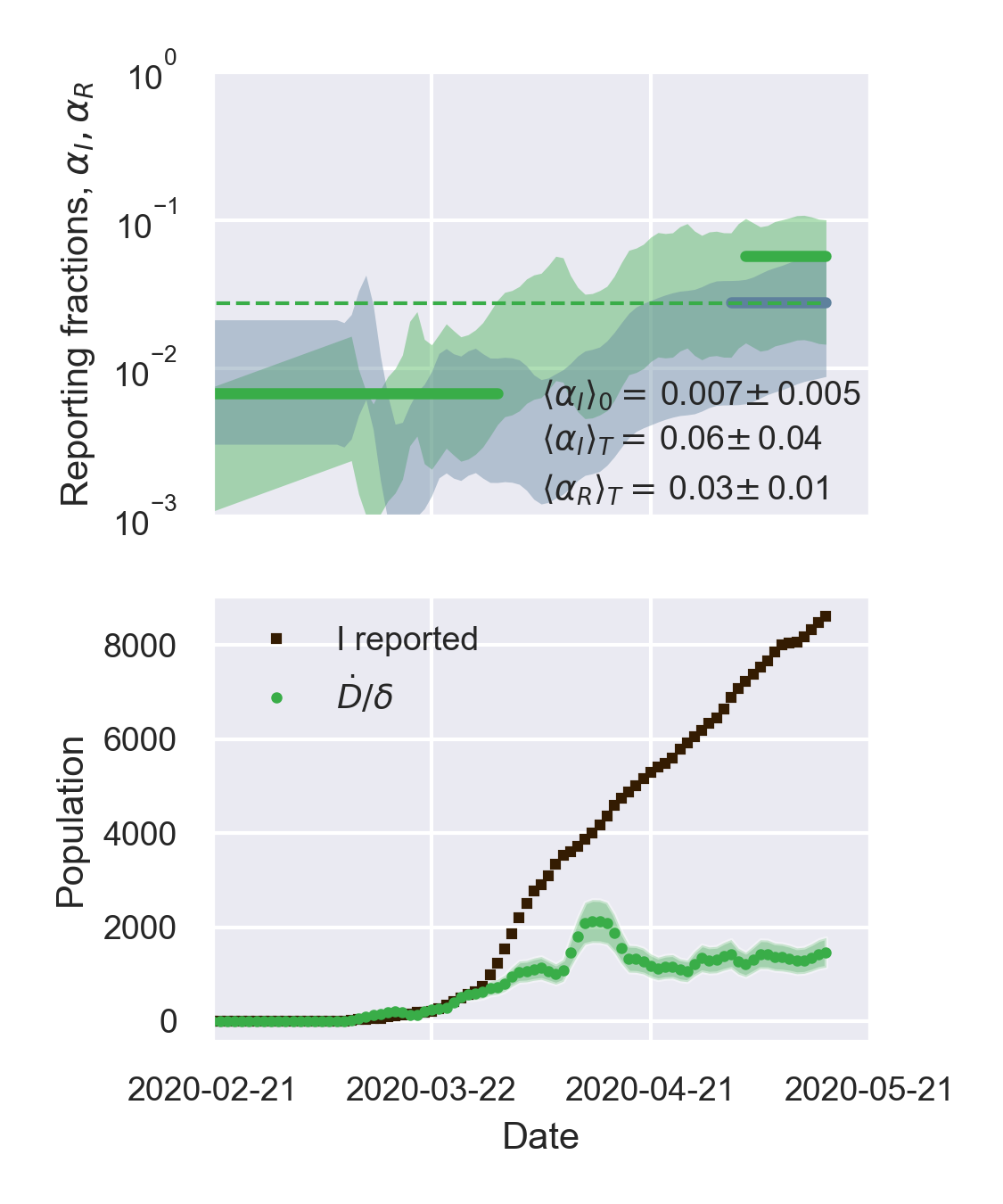} 
\end{tabular}
\caption{\textbf{A positive derivative of the reporting fraction of infected causes a delay of
the apparent epidemic peak with respect to the true one - II.}
Top panels: fraction of true recovered reported, $\alpha_R(t)$ (grey shaded regions) and 
fraction of true infected reported, $\alpha_I(t)$, computed as described in the text. The shaded
regions correspond to an assumed mortality $m = 0.012 \pm 0.005$~\cite{Bastolla:2020aa}. 
The thick lines mark averages on the last 15 \% portions of the data ($\langle\alpha_{I,R}\rangle_T$), 
and first 40 \% ($\langle\alpha_I\rangle_0$), while the dashed lines
denote the overall average values of $\alpha_{I,R}$ computed over the whole available time span.
Bottom panels. The reported infected are compared to the rescaled deaths/day, 
$I_0(t)~\stackrel{\rm def}{=}~\dot{D}\alpha_I(0)/d$, i.e. the 
infected that would have been reported if the reporting fraction had 
remained constant at its initial value $\alpha_I(0)$. Technically, $\delta = d/\alpha_I(0)$ is 
obtained as the slope $\delta$ of a linear fit of $\dot{D}$ vs $I_M$ in the very early stage of the outbreak.
The shaded regions corresponds to the least-square errors found on $\delta$.
}
\label{fig:alphadot12}
\end{figure}

\begin{figure}[t!]
\centering
\begin{tabular}{ccc}
Egypt & Morocco & Algeria \\
\includegraphics [width=0.33\textwidth]{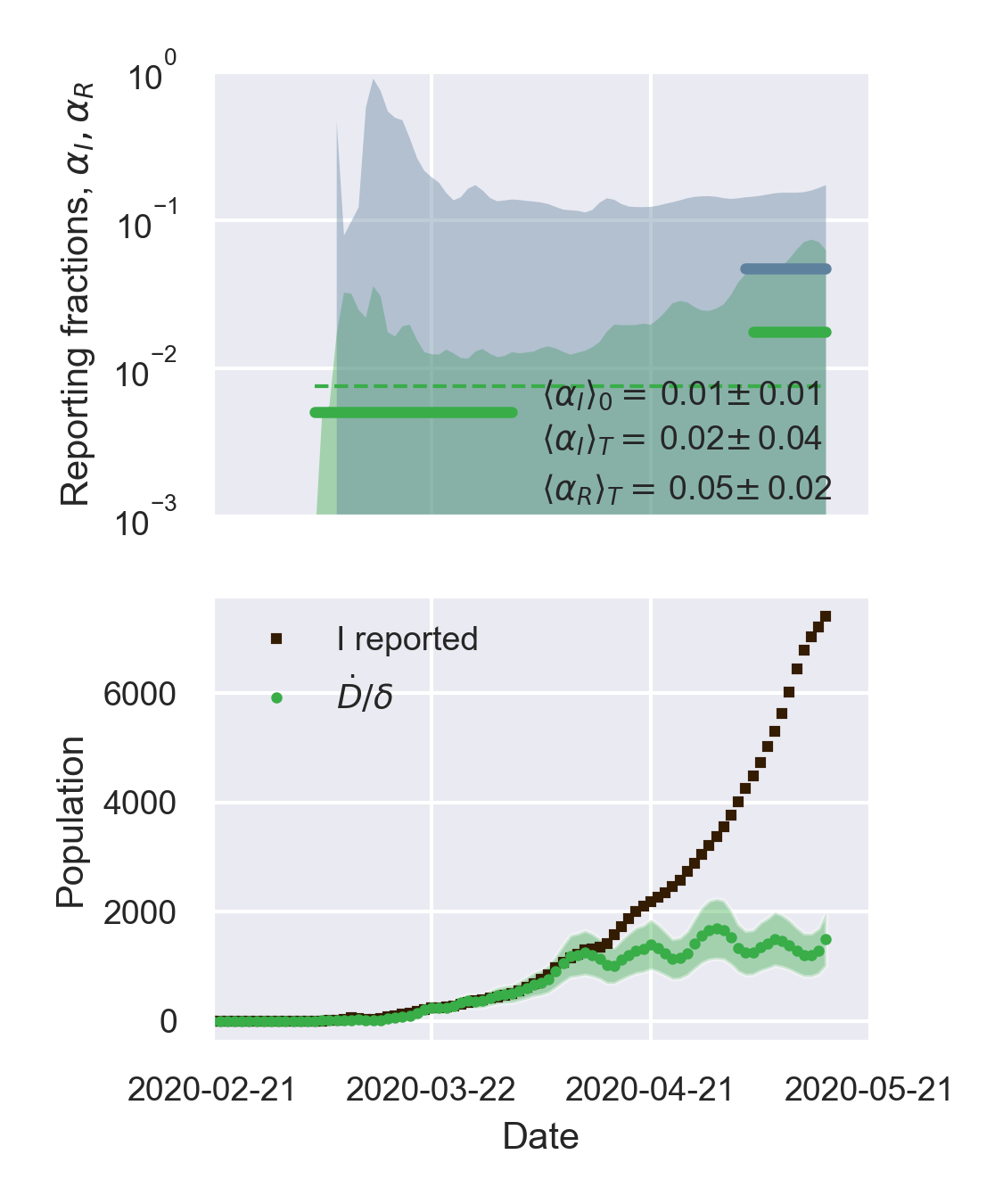} &  
\includegraphics [width=0.33\textwidth]{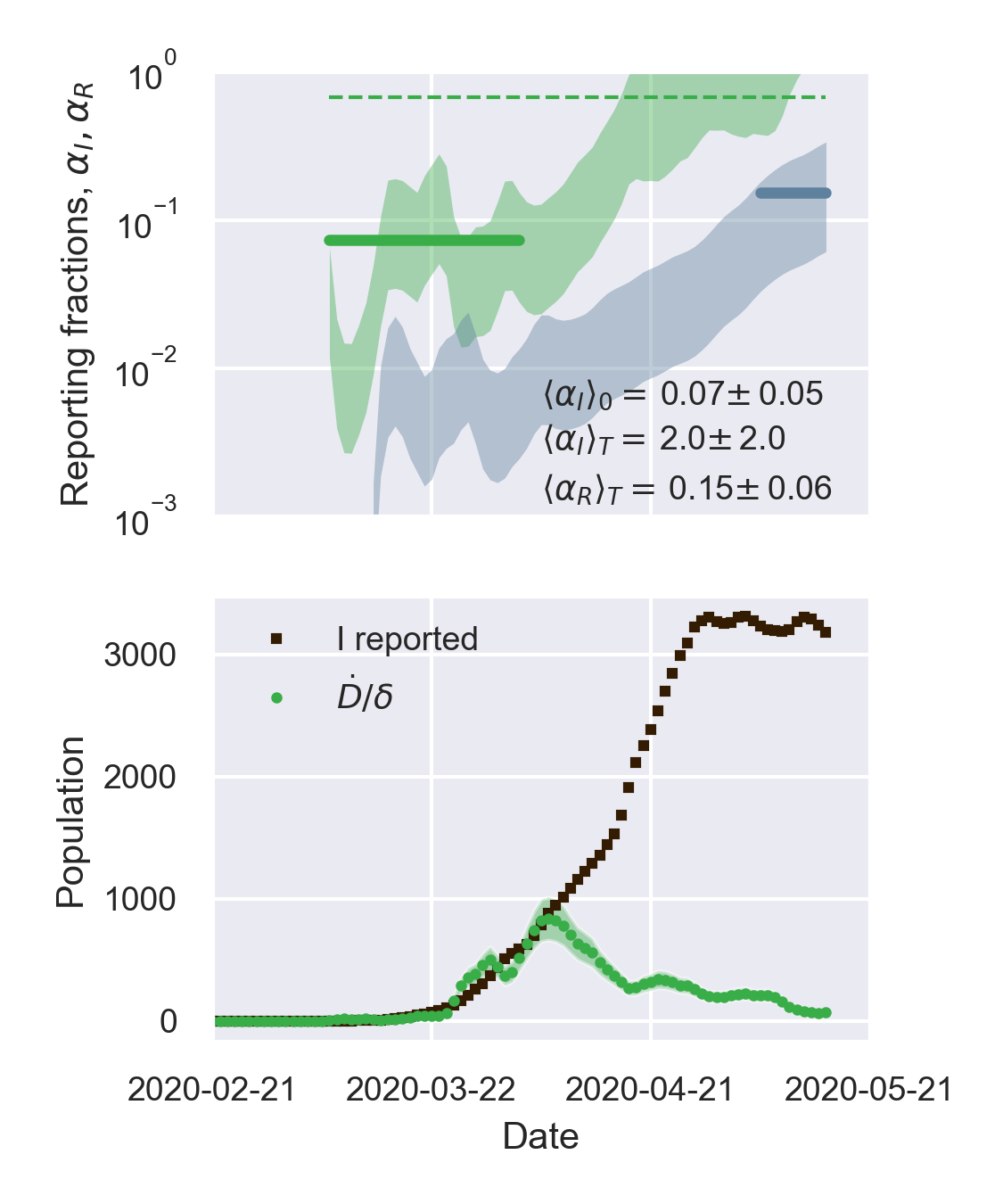} & 
\includegraphics [width=0.33\textwidth]{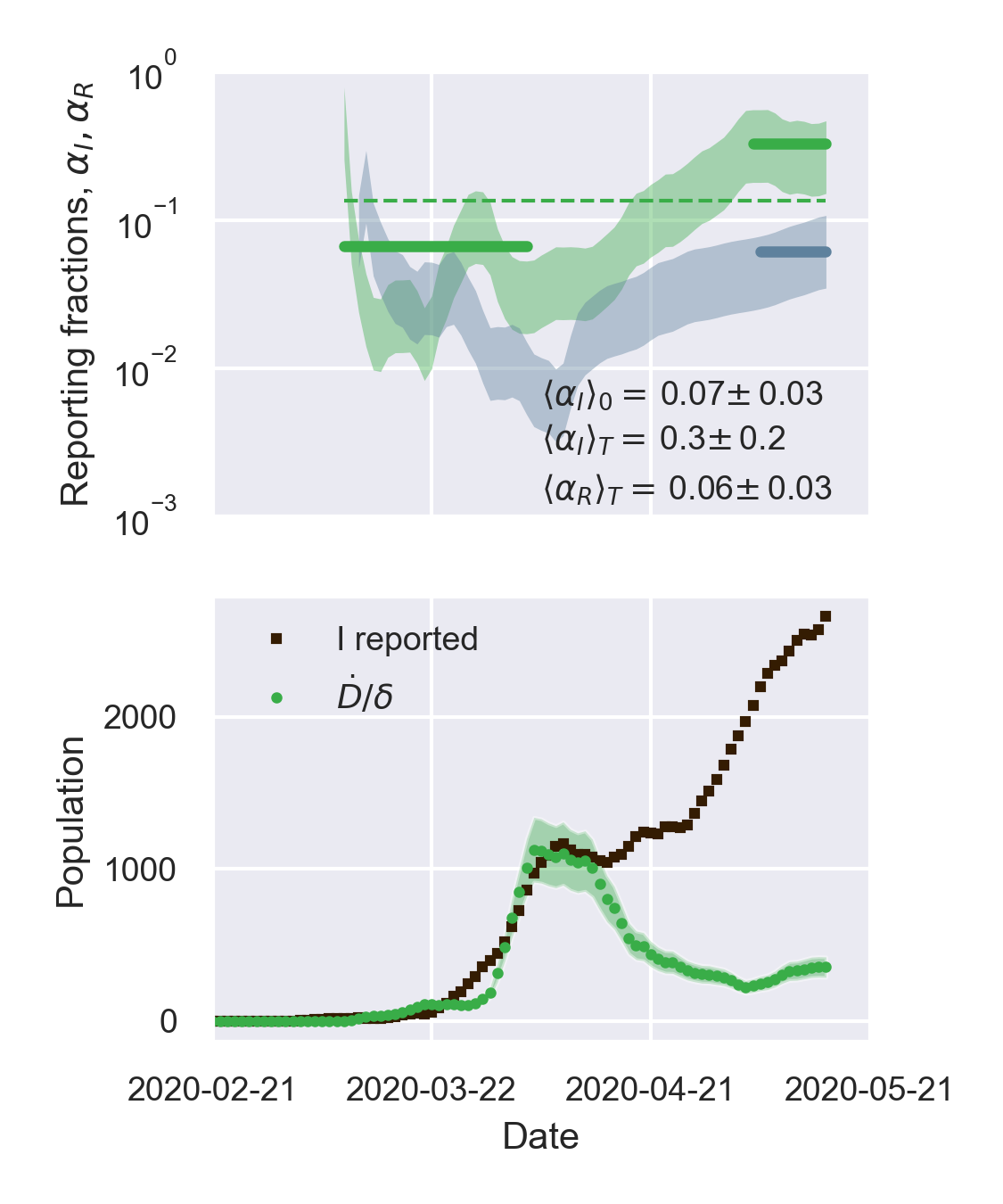}
\end{tabular}
\caption{\textbf{A positive derivative of the reporting fraction of infected causes a delay of
the apparent epidemic peak with respect to the true one - III.}
Top panels: fraction of true recovered reported, $\alpha_R(t)$ (grey shaded regions) and 
fraction of true infected reported, $\alpha_I(t)$, computed as described in the text. The shaded
regions correspond to an assumed mortality $m = 0.012 \pm 0.005$~\cite{Bastolla:2020aa}. 
The thick lines mark averages on the last 15 \% portions of the data ($\langle\alpha_{I,R}\rangle_T$), 
and first 40 \% ($\langle\alpha_I\rangle_0$), while the dashed lines
denote the overall average values of $\alpha_{I,R}$ computed over the whole available time span.
Bottom panels. The reported infected are compared to the rescaled deaths/day, 
$I_0(t)~\stackrel{\rm def}{=}~\dot{D}\alpha_I(0)/d$, i.e. the 
infected that would have been reported if the reporting fraction had 
remained constant at its initial value $\alpha_I(0)$. Technically, $\delta = d/\alpha_I(0)$ is 
obtained as the slope $\delta$ of a linear fit of $\dot{D}$ vs $I_M$ in the very early stage of the outbreak.
The shaded regions corresponds to the least-square errors found on $\delta$.
}
\label{fig:alphadot13}
\end{figure}
 
\begin{figure}[t!]
\centering
\begin{tabular}{ccc}
France & Switzerland  & Austria \\
\includegraphics [width=0.33\textwidth]{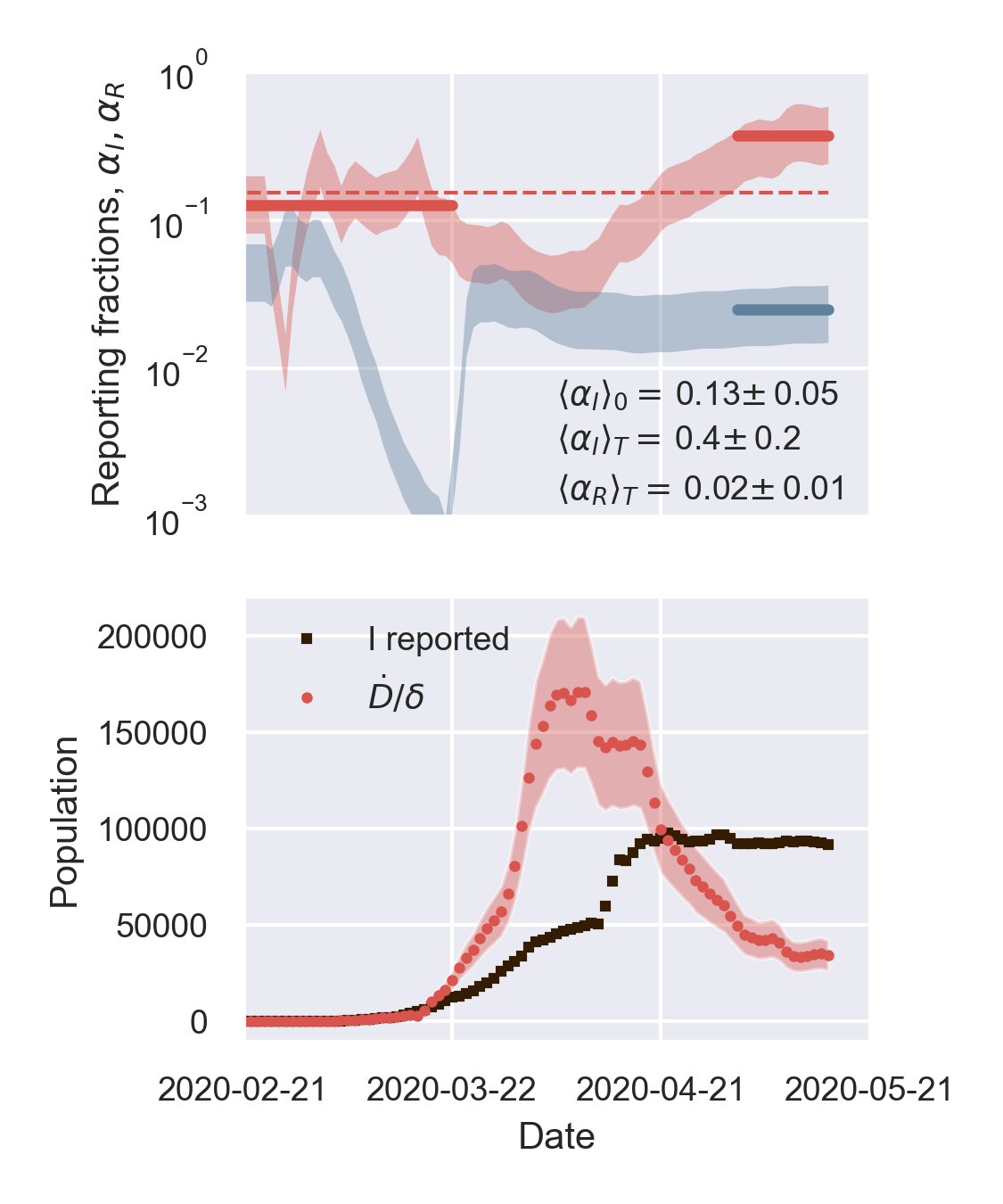}  &  
\includegraphics [width=0.33\textwidth]{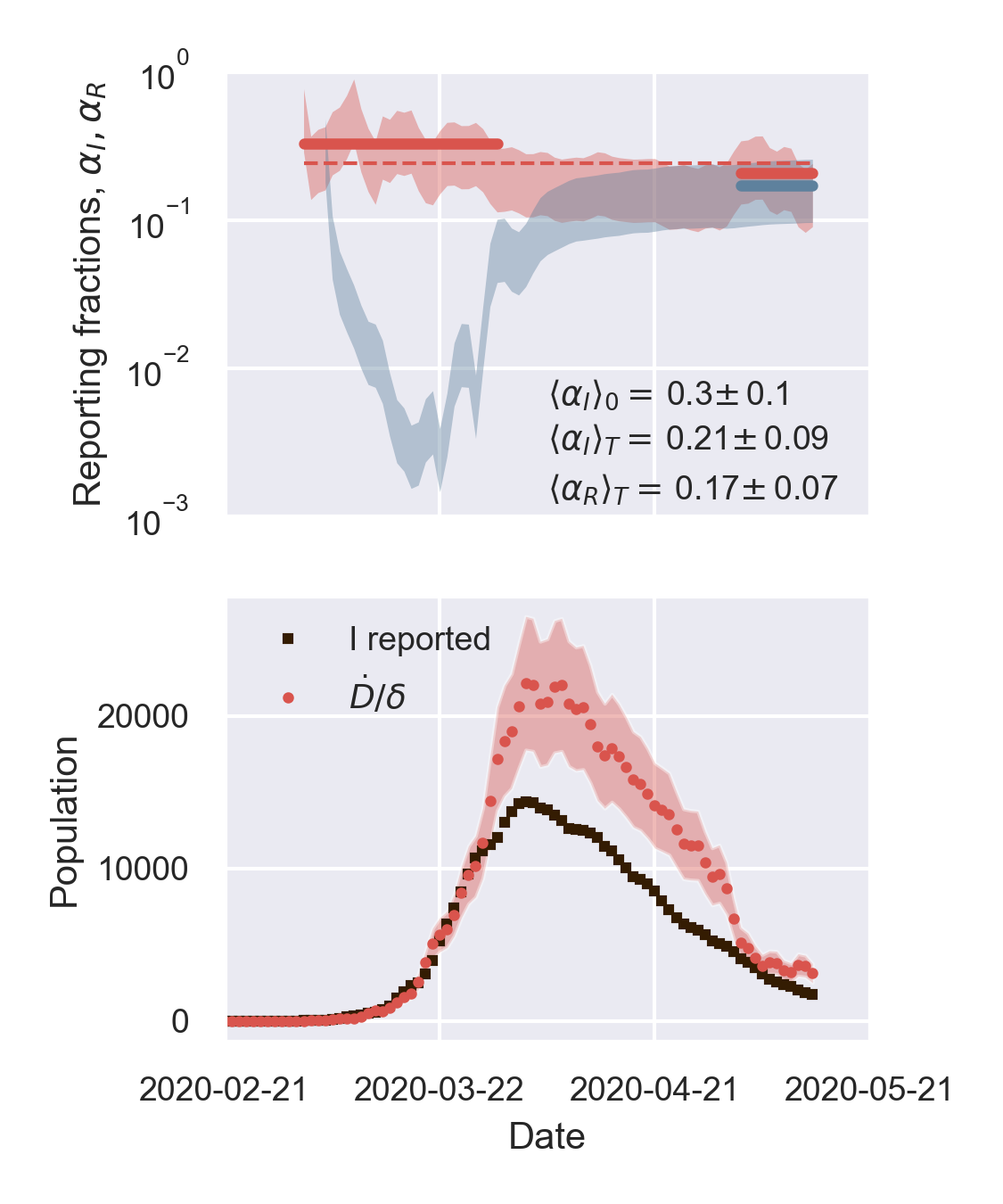} & 
\includegraphics [width=0.33\textwidth]{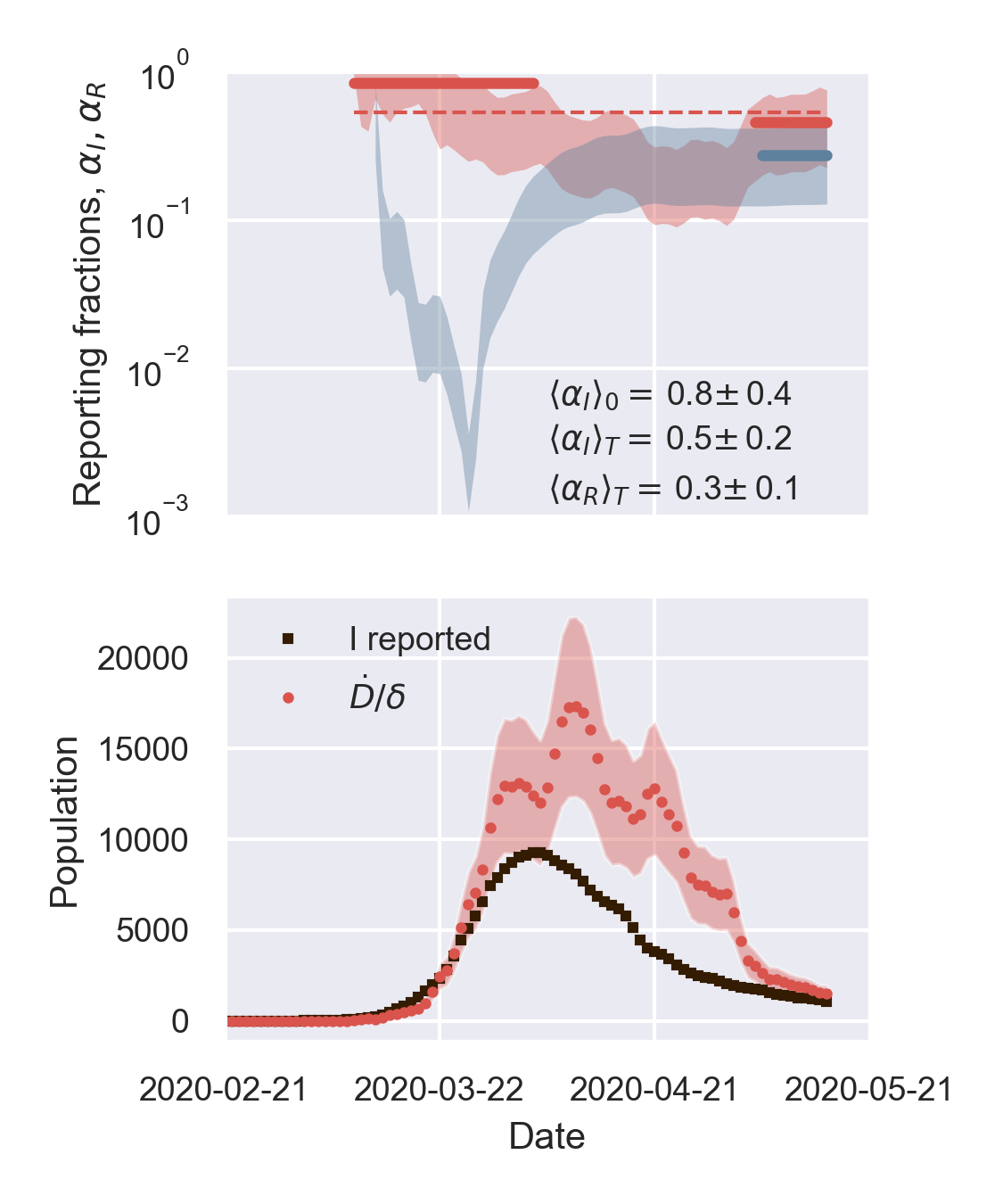} \\
Finland &  Canada &  \\
\includegraphics [width=0.33\textwidth]{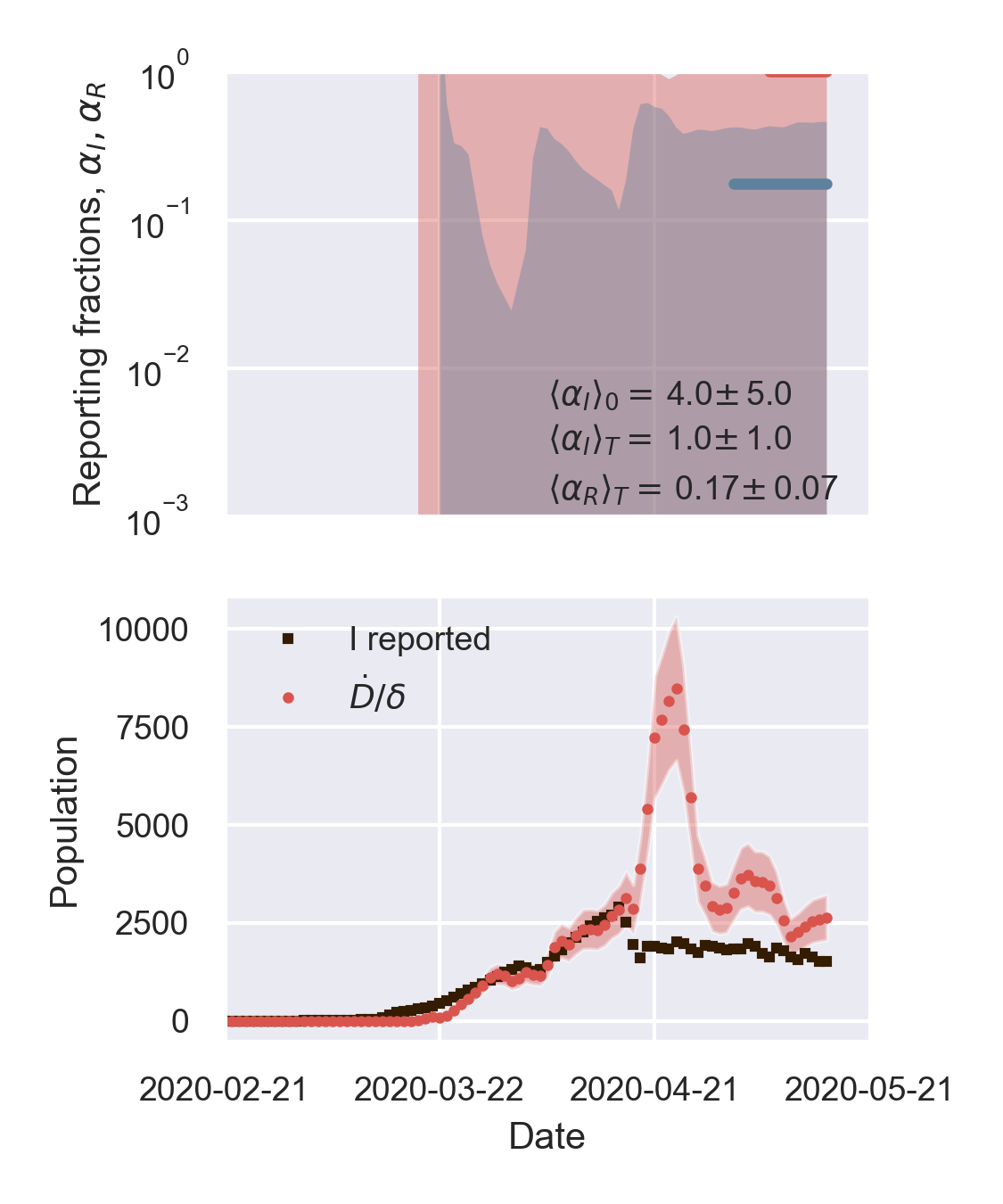}  &  
\includegraphics [width=0.33\textwidth]{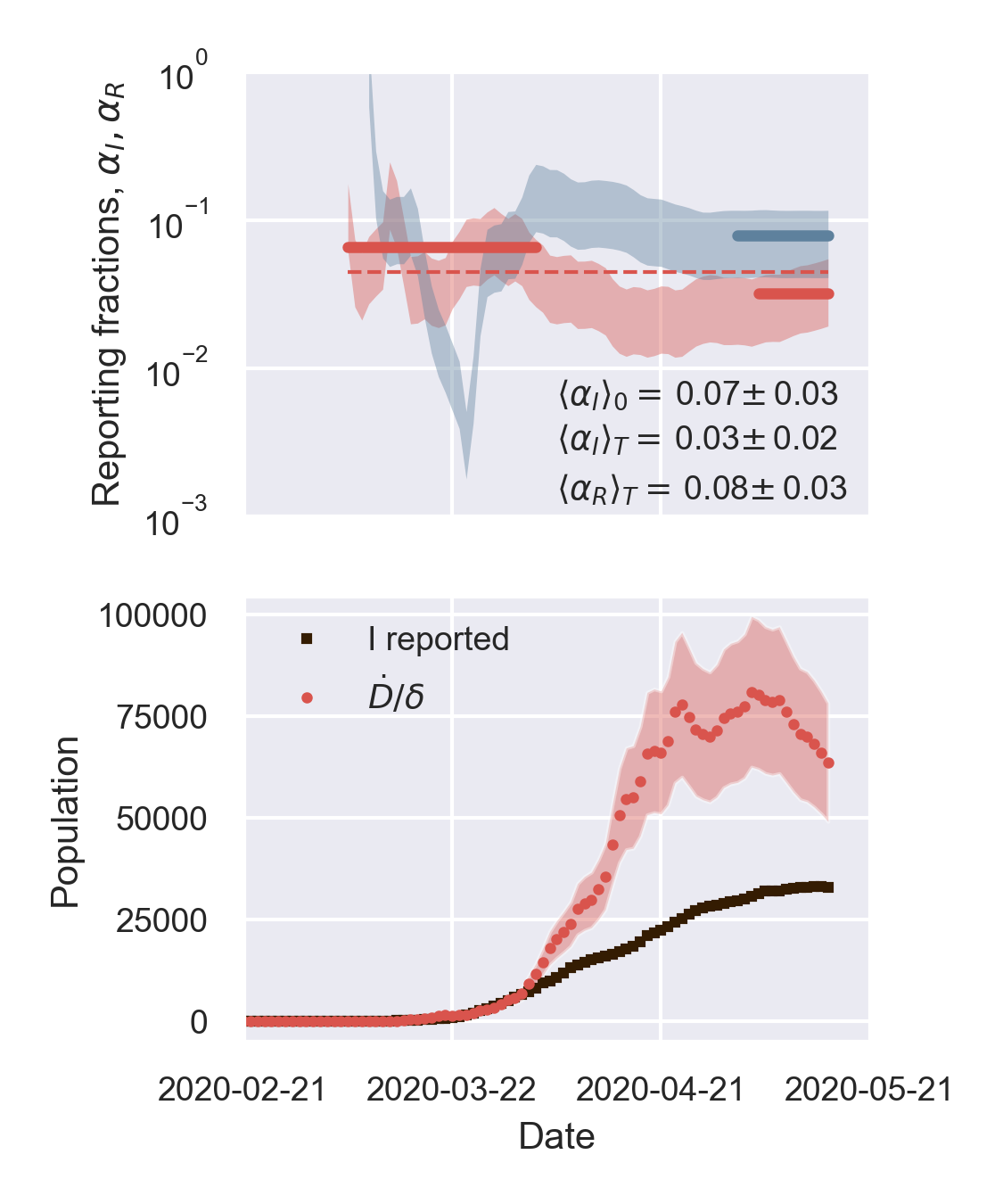} & 
\end{tabular}
\caption{\textbf{A negative derivative of the reporting fraction of infected deceptively anticipates
the apparent epidemic peak with respect to the true one.}
Top panels: fraction of true recovered reported, $\alpha_R(t)$ (grey shaded regions) and 
fraction of true infected reported, $\alpha_I(t)$, computed as described in the text. The shaded
regions correspond to an assumed mortality $m = 0.012 \pm 0.005$~\cite{Bastolla:2020aa}. 
The thick lines mark averages on the last 15 \% portions of the data ($\langle\alpha_{I,R}\rangle_T$), 
and first 40 \% ($\langle\alpha_I\rangle_0$), while the dashed lines
denote the overall average values of $\alpha_{I,R}$ computed over the whole available time span.
Bottom panels. The reported infected are compared to the rescaled deaths/day, 
$I_0(t)~\stackrel{\rm def}{=}~\dot{D}\alpha_I(0)/d$, i.e. the 
infected that would have been reported if the reporting fraction had 
remained constant at its initial value $\alpha_I(0)$. Technically, $\delta = d/\alpha_I(0)$ is 
obtained as the slope $\delta$ of a linear fit of $\dot{D}$ vs $I_M$ in the very early stage of the outbreak.
The shaded regions corresponds to the least-square errors found on $\delta$.
}
\label{fig:alphadot2}
\end{figure}

\begin{figure}[t!]
\centering
\begin{tabular}{cc}
Israel  & Denmark \\
\includegraphics [width=0.5\textwidth]{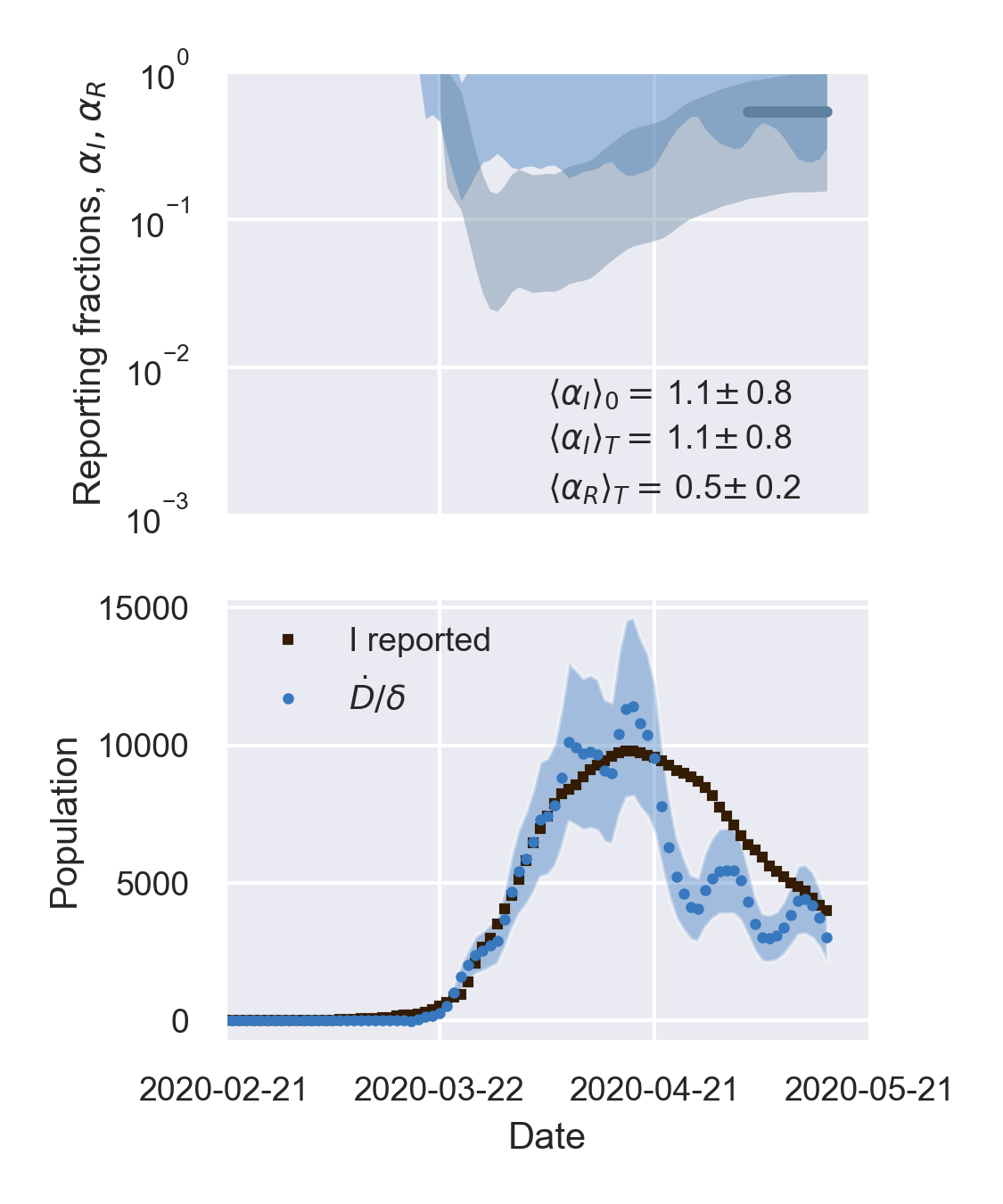}  &  
\includegraphics [width=0.5\textwidth]{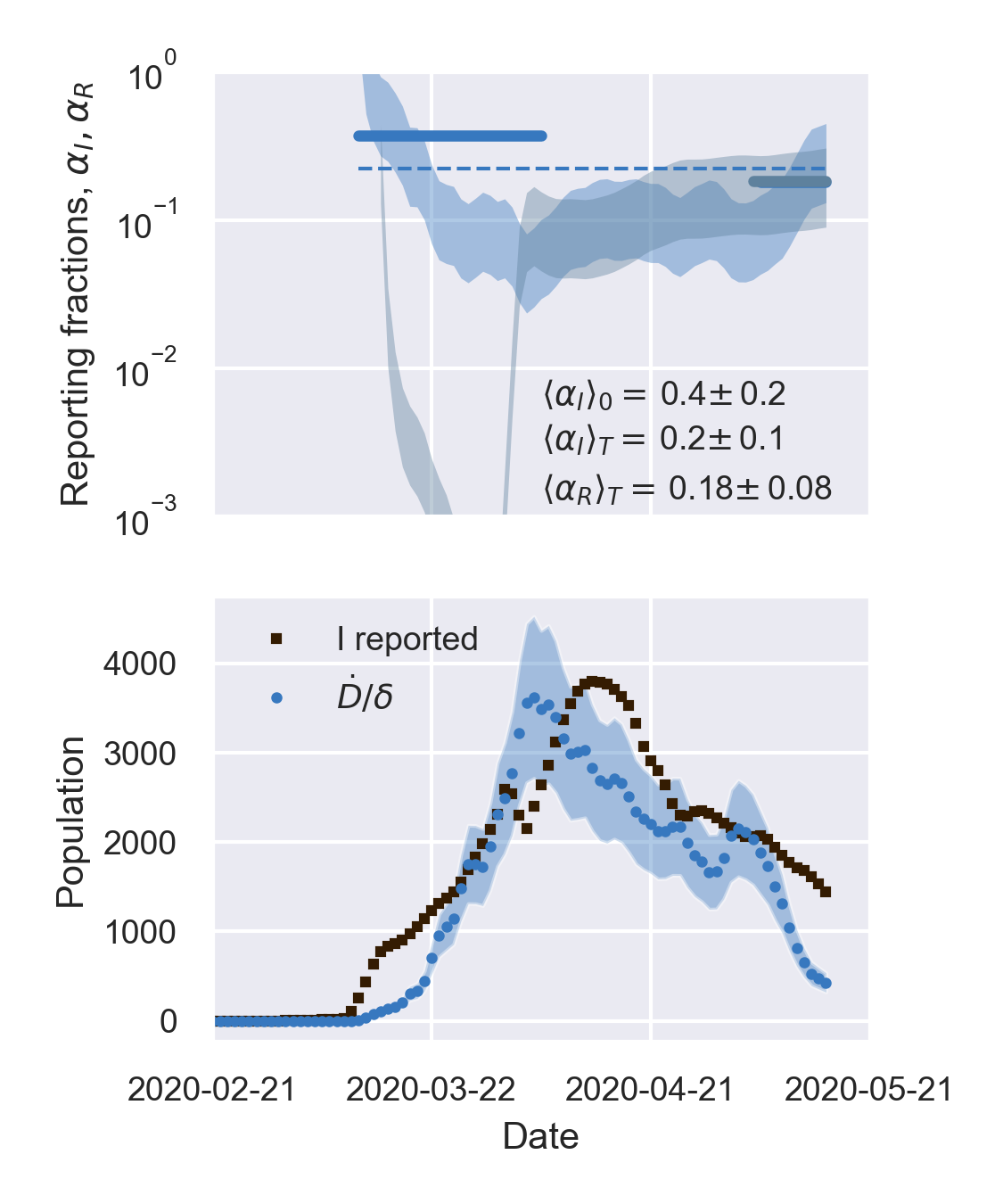} 
\end{tabular}
\caption{\textbf{The outbreak in a reduced number of {\em marginal} cases 
is consistent with an
overall flat  reporting fraction of infected.}
Top panels: fraction of true recovered reported, $\alpha_R(t)$ (grey shaded regions) and 
fraction of true infected reported, $\alpha_I(t)$, computed as described in the text. The shaded
regions correspond to an assumed mortality $m = 0.012 \pm 0.005$~\cite{Bastolla:2020aa}. 
The thick lines mark averages on the last 15 \% portions of the data ($\langle\alpha_{I,R}\rangle_T$), 
and first 40 \% ($\langle\alpha_I\rangle_0$), while the dashed lines
denote the overall average values of $\alpha_{I,R}$ computed over the whole available time span.
Bottom panels. The reported infected are compared to the rescaled deaths/day, 
$I_0(t)~\stackrel{\rm def}{=}~\dot{D}\alpha_I(0)/d$, i.e. the 
infected that would have been reported if the reporting fraction had 
remained constant at its initial value $\alpha_I(0)$. Technically, $\delta = d/\alpha_I(0)$ is 
obtained as the slope $\delta$ of a linear fit of $\dot{D}$ vs $I_M$ in the very early stage of the outbreak.
The shaded regions corresponds to the least-square errors found on $\delta$.
}
\label{fig:alphadot3}
\end{figure}

\clearpage

%
%

\providecommand{\refin}[1]{\\ \textbf{Referenced in:} #1}
\begin{thebibliography}{10}

\bibitem{achemso-control}


\bibitem{Mena2000}
Abarca,~A.;\ \ G{\'o}mez-Sal,~P.;\ \ Mart{\'i}n,~A.;\ \ Mena,~M.;\ \
  Poblet,~J.~M.;\ \ Y{\'e}lamos,~C. \textit{Inorg. Chem.} \textbf{2000,}
  \textsl{39,} 642--651.

\bibitem{Abernethy2003}
Abernethy,~C.~D.;\ \ Codd,~G.~M.;\ \ Spicer,~M.~D.;\ \ Taylor,~M.~K.
  \textit{J.~Am. Chem. Soc.} \textbf{2003,} \textsl{125,} 1128--1129.

\bibitem{Friedman-Hill2003}
Friedman-Hill,~E.  Writing Rules in Jess.   In  \textit{{Jess} in Action:
  {Java} Rule-based Systems}, 1 ed.;\ Manning Publications Co.: Greenwich, CT,
  USA, 2003.

\bibitem{EuropeanCommission2008}
``Communication from the European Commission to the European Council and the
  European Parliament: 20 20 by 2020: Europe's climate change opportunity'',
  Technical Report,  Brussels, Belgium, 2008.

\bibitem{Cotton1999}
Cotton,~F.~A.;\ \ Wilkinson,~G.;\ \ Murillio,~C.~A.;\ \ Bochmann,~M.
  \textit{Advanced Inorganic Chemistry;} Wiley: Chichester, United Kingdom, 6
  ed.; 1999.

\bibitem{Pople2003}
Frisch,~M.~J. \textit{et al.}\  ``{Gaussian} 03'',  Gaussian, Inc.,
  Wallingford, CT, 2004.

\bibitem{Johnson1972}
Johnson,~A.~L. ``1-(Alkylsubstituted phenyl)imidazoles useful in {ACTH} reverse
  assay'',  1972.

\bibitem{Arduengo1992}
Arduengo,~III,~A.~J.;\ \ Dias,~H. V.~R.;\ \ Harlow,~R.~L.;\ \ Kline,~M.
  \textit{J.~Am.\ Chem.\ Soc.} \textbf{1992,} \textsl{114,} 5530--5534.

\bibitem{Eisenstein2005}
Appelhans,~L.~N.;\ \ Zuccaccia,~D.;\ \ Kovacevic,~A.;\ \ Chianese,~A.~R.;\ \
  Miecznikowski,~J.~R.;\ \ Macchioni,~A.;\ \ Clot,~E.;\ \ Eisenstein,~O.;\ \
  Crabtree,~R.~H. \textit{J.~Am.\ Chem. Soc.} \textbf{2005,} \textsl{127,}
  16299--16311.

\bibitem{Arduengo1994}
Arduengo,~III,~A.~J.;\ \ Gamper,~S.~F.;\ \ Calabrese,~J.~C.;\ \ Davidson,~F.
  \textbf{1994,} \textsl{116,} 4391--4394.

\bibitem{Note-1}
This is a note. The text will be moved the the references section. The title of
  the section will change to ``Notes and References''.

\bibitem{Note-2}
Ref.~\citenum {Cotton1999}, p.~1.

\end{thebibliography}


\begin{mcitethebibliography}{24}
\providecommand*\natexlab[1]{#1}
\providecommand*\mciteSetBstSublistMode[1]{}
\providecommand*\mciteSetBstMaxWidthForm[2]{}
\providecommand*\mciteBstWouldAddEndPuncttrue
  {\def\EndOfBibitem{\unskip.}}
\providecommand*\mciteBstWouldAddEndPunctfalse
  {\let\EndOfBibitem\relax}
\providecommand*\mciteSetBstMidEndSepPunct[3]{}
\providecommand*\mciteSetBstSublistLabelBeginEnd[3]{}
\providecommand*\EndOfBibitem{}
\mciteSetBstSublistMode{f}
\mciteSetBstMaxWidthForm{subitem}{(\alph{mcitesubitemcount})}
\mciteSetBstSublistLabelBeginEnd
  {\mcitemaxwidthsubitemform\space}
  {\relax}
  {\relax}

\bibitem[WHO(2020)]{WHO:2020aa}
2020;
  \url{https://www.who.int/emergencies/diseases/novel-coronavirus-2019}\relax
\mciteBstWouldAddEndPuncttrue
\mciteSetBstMidEndSepPunct{\mcitedefaultmidpunct}
{\mcitedefaultendpunct}{\mcitedefaultseppunct}\relax
\EndOfBibitem
\bibitem[Layne \latin{et~al.}(2020)Layne, Hyman, Morens, and
  Taubenberger]{Layne:2020aa}
Layne,~S.~P.; Hyman,~J.~M.; Morens,~D.~M.; Taubenberger,~J.~K. {New coronavirus
  outbreak: Framing questions for pandemic prevention}. \emph{Science
  Translational Medicine} \textbf{2020}, \emph{12}, eabb1469\relax
\mciteBstWouldAddEndPuncttrue
\mciteSetBstMidEndSepPunct{\mcitedefaultmidpunct}
{\mcitedefaultendpunct}{\mcitedefaultseppunct}\relax
\EndOfBibitem
\bibitem[Enserink and Kupferschmidt(2020)Enserink, and
  Kupferschmidt]{Enserink:2020aa}
Enserink,~M.; Kupferschmidt,~K. {With COVID-19, modeling takes on life and
  death importance}. \emph{Science} \textbf{2020}, \emph{367}, 1414--1415\relax
\mciteBstWouldAddEndPuncttrue
\mciteSetBstMidEndSepPunct{\mcitedefaultmidpunct}
{\mcitedefaultendpunct}{\mcitedefaultseppunct}\relax
\EndOfBibitem
\bibitem[Wang \latin{et~al.}(2020)Wang, Wang, Chen, and Qin]{Wang:2020aa}
Wang,~Y.; Wang,~Y.; Chen,~Y.; Qin,~Q. {Unique epidemiological and clinical
  features of the emerging 2019 novel coronavirus pneumonia (COVID-19)
  implicate special control measures}. \emph{Journal of Medical Virology}
  \textbf{2020}, \emph{92}, 568--576\relax
\mciteBstWouldAddEndPuncttrue
\mciteSetBstMidEndSepPunct{\mcitedefaultmidpunct}
{\mcitedefaultendpunct}{\mcitedefaultseppunct}\relax
\EndOfBibitem
\bibitem[Kupferschmidt(2020)]{Kupferschmidt:2020aa}
Kupferschmidt,~K. Why do some {COVID}-19 patients infect many others, whereas
  most don't spread the virus at all? \emph{Science} \textbf{2020}, \relax
\mciteBstWouldAddEndPunctfalse
\mciteSetBstMidEndSepPunct{\mcitedefaultmidpunct}
{}{\mcitedefaultseppunct}\relax
\EndOfBibitem
\bibitem[Gatto \latin{et~al.}(2020)Gatto, Bertuzzo, Mari, Miccoli, Carraro,
  Casagrandi, and Rinaldo]{Gatto:2020aa}
Gatto,~M.; Bertuzzo,~E.; Mari,~L.; Miccoli,~S.; Carraro,~L.; Casagrandi,~R.;
  Rinaldo,~A. {Spread and dynamics of the COVID-19 epidemic in Italy: Effects
  of emergency containment measures}. \emph{Proceedings of the National Academy
  of Sciences} \textbf{2020}, \emph{117}, 10484--10491\relax
\mciteBstWouldAddEndPuncttrue
\mciteSetBstMidEndSepPunct{\mcitedefaultmidpunct}
{\mcitedefaultendpunct}{\mcitedefaultseppunct}\relax
\EndOfBibitem
\bibitem[Giordano \latin{et~al.}(2020)Giordano, Blanchini, Bruno, Colaneri,
  Di~Filippo, Di~Matteo, and Colaneri]{Giordano:2020aa}
Giordano,~G.; Blanchini,~F.; Bruno,~R.; Colaneri,~P.; Di~Filippo,~A.;
  Di~Matteo,~A.; Colaneri,~M. Modelling the COVID-19 epidemic and
  implementation of population-wide interventions in Italy. \emph{Nature
  Medicine} \textbf{2020}, \relax
\mciteBstWouldAddEndPunctfalse
\mciteSetBstMidEndSepPunct{\mcitedefaultmidpunct}
{}{\mcitedefaultseppunct}\relax
\EndOfBibitem
\bibitem[Chinazzi \latin{et~al.}(2020)Chinazzi, Davis, Ajelli, Gioannini,
  Litvinova, Merler, {Pastore Y Piontti}, Mu, Rossi, Sun, Viboud, Xiong, Yu,
  Halloran, Longini, and Vespignani]{Chinazzi:2020aa}
Chinazzi,~M. \latin{et~al.}  {The effect of travel restrictions on the spread
  of the 2019 novel coronavirus (COVID-19) outbreak}. \emph{Science (New York,
  N.Y.)} \textbf{2020}, \emph{368}, 395--400\relax
\mciteBstWouldAddEndPuncttrue
\mciteSetBstMidEndSepPunct{\mcitedefaultmidpunct}
{\mcitedefaultendpunct}{\mcitedefaultseppunct}\relax
\EndOfBibitem
\bibitem[Balcan \latin{et~al.}(2009)Balcan, Colizza, Gon{\c c}alves, Hu,
  Ramasco, and Vespignani]{Balcan:2009aa}
Balcan,~D.; Colizza,~V.; Gon{\c c}alves,~B.; Hu,~H.; Ramasco,~J.~J.;
  Vespignani,~A. Multiscale mobility networks and the spatial spreading of
  infectious diseases. \emph{Proceedings of the National Academy of Sciences}
  \textbf{2009}, \emph{106}, 21484--21489\relax
\mciteBstWouldAddEndPuncttrue
\mciteSetBstMidEndSepPunct{\mcitedefaultmidpunct}
{\mcitedefaultendpunct}{\mcitedefaultseppunct}\relax
\EndOfBibitem
\bibitem[Roda \latin{et~al.}(2020)Roda, Varughese, Han, and Li]{Roda:2020aa}
Roda,~W.~C.; Varughese,~M.~B.; Han,~D.; Li,~M.~Y. Why is it difficult to
  accurately predict the COVID-19 epidemic? \emph{Infectious Disease Modelling}
  \textbf{2020}, \emph{5}, 271 -- 281\relax
\mciteBstWouldAddEndPuncttrue
\mciteSetBstMidEndSepPunct{\mcitedefaultmidpunct}
{\mcitedefaultendpunct}{\mcitedefaultseppunct}\relax
\EndOfBibitem
\bibitem[De~Brouwer \latin{et~al.}(2020)De~Brouwer, Raimondi, and
  Moreau]{De-Brouwer:2020aa}
De~Brouwer,~E.; Raimondi,~D.; Moreau,~Y. Modeling the COVID-19 outbreaks and
  the effectiveness of the containment measures adopted across countries.
  \emph{medRxiv} \textbf{2020}, 2020.04.02.20046375\relax
\mciteBstWouldAddEndPuncttrue
\mciteSetBstMidEndSepPunct{\mcitedefaultmidpunct}
{\mcitedefaultendpunct}{\mcitedefaultseppunct}\relax
\EndOfBibitem
\bibitem[Hitchcock(2020)]{Hitchcock:2020aa}
Hitchcock,~C. In \emph{The Stanford Encyclopedia of Philosophy}, summer 2020
  ed.; Zalta,~E.~N., Ed.; Metaphysics Research Lab, Stanford University,
  2020\relax
\mciteBstWouldAddEndPuncttrue
\mciteSetBstMidEndSepPunct{\mcitedefaultmidpunct}
{\mcitedefaultendpunct}{\mcitedefaultseppunct}\relax
\EndOfBibitem
\bibitem[Kermack \latin{et~al.}(1927)Kermack, McKendrick, and
  Walker]{Kermack:1927aa}
Kermack,~W.~O.; McKendrick,~A.~G.; Walker,~G.~T. A contribution to the
  mathematical theory of epidemics. \emph{Proceedings of the Royal Society of
  London. Series A, Containing Papers of a Mathematical and Physical Character}
  \textbf{1927}, \emph{115}, 700--721\relax
\mciteBstWouldAddEndPuncttrue
\mciteSetBstMidEndSepPunct{\mcitedefaultmidpunct}
{\mcitedefaultendpunct}{\mcitedefaultseppunct}\relax
\EndOfBibitem
\bibitem[Fanelli and Piazza(2020)Fanelli, and Piazza]{Fanelli:2020aa}
Fanelli,~D.; Piazza,~F. Analysis and forecast of COVID-19 spreading in China,
  Italy and France. \emph{Chaos, Solitons \& Fractals} \textbf{2020},
  \emph{134}, 109761\relax
\mciteBstWouldAddEndPuncttrue
\mciteSetBstMidEndSepPunct{\mcitedefaultmidpunct}
{\mcitedefaultendpunct}{\mcitedefaultseppunct}\relax
\EndOfBibitem
\bibitem[Anastassopoulou \latin{et~al.}(2020)Anastassopoulou, Russo, Tsakris,
  and Siettos]{Anastassopoulou:2020aa}
Anastassopoulou,~C.; Russo,~L.; Tsakris,~A.; Siettos,~C. {Data-based analysis,
  modelling and forecasting of the COVID-19 outbreak}. \emph{PLoS ONE}
  \textbf{2020}, \emph{15}, e0230405\relax
\mciteBstWouldAddEndPuncttrue
\mciteSetBstMidEndSepPunct{\mcitedefaultmidpunct}
{\mcitedefaultendpunct}{\mcitedefaultseppunct}\relax
\EndOfBibitem
\bibitem[Murray(1989)]{Murray:1989aa}
Murray,~J.~D. \emph{Mathematical Biology}; Springer, 1989\relax
\mciteBstWouldAddEndPuncttrue
\mciteSetBstMidEndSepPunct{\mcitedefaultmidpunct}
{\mcitedefaultendpunct}{\mcitedefaultseppunct}\relax
\EndOfBibitem
\bibitem[Vattay(2020)]{Vattay:2020aa}
Vattay,~G. Forecasting the outcome and estimating the epidemic model parameters
  from the fatality time series in COVID-19 outbreaks. 2020\relax
\mciteBstWouldAddEndPuncttrue
\mciteSetBstMidEndSepPunct{\mcitedefaultmidpunct}
{\mcitedefaultendpunct}{\mcitedefaultseppunct}\relax
\EndOfBibitem
\bibitem[Dehning \latin{et~al.}(2020)Dehning, Zierenberg, Spitzner, Wibral,
  Neto, Wilczek, and Priesemann]{Dehning:2020aa}
Dehning,~J.; Zierenberg,~J.; Spitzner,~F.~P.; Wibral,~M.; Neto,~J.~P.;
  Wilczek,~M.; Priesemann,~V. Inferring change points in the spread of COVID-19
  reveals the effectiveness of interventions. \emph{Science} \textbf{2020},
  \relax
\mciteBstWouldAddEndPunctfalse
\mciteSetBstMidEndSepPunct{\mcitedefaultmidpunct}
{}{\mcitedefaultseppunct}\relax
\EndOfBibitem
\bibitem[Brauer and Castillo-Chavez(2012)Brauer, and
  Castillo-Chavez]{Brauer:2012aa}
Brauer,~F.; Castillo-Chavez,~C. \emph{Mathematical Models in Population Biology
  and Epidemiology}, 2nd ed.; Springer, 2012\relax
\mciteBstWouldAddEndPuncttrue
\mciteSetBstMidEndSepPunct{\mcitedefaultmidpunct}
{\mcitedefaultendpunct}{\mcitedefaultseppunct}\relax
\EndOfBibitem
\bibitem[Diekmann and Heesterbeek(2000)Diekmann, and
  Heesterbeek]{Diekmann:2000aa}
Diekmann,~O.; Heesterbeek,~J. A.~P. \emph{Mathematical Epidemiology of
  Infectious Diseases: Model Building, Analysis and Interpretation}; Wiley,
  2000\relax
\mciteBstWouldAddEndPuncttrue
\mciteSetBstMidEndSepPunct{\mcitedefaultmidpunct}
{\mcitedefaultendpunct}{\mcitedefaultseppunct}\relax
\EndOfBibitem
\bibitem[Bastolla(2020)]{Bastolla:2020aa}
Bastolla,~U. How lethal is the novel coronavirus, and how many undetected cases
  there are? The importance of being tested. \emph{medRxiv} \textbf{2020},
  \relax
\mciteBstWouldAddEndPunctfalse
\mciteSetBstMidEndSepPunct{\mcitedefaultmidpunct}
{}{\mcitedefaultseppunct}\relax
\EndOfBibitem
\bibitem[Peto(2020)]{Peto:2020aa}
Peto,~J. {Covid-19 mass testing facilities could end the epidemic rapidly}.
  \emph{British Medical Journal} \textbf{2020}, \emph{368}, m1163\relax
\mciteBstWouldAddEndPuncttrue
\mciteSetBstMidEndSepPunct{\mcitedefaultmidpunct}
{\mcitedefaultendpunct}{\mcitedefaultseppunct}\relax
\EndOfBibitem
\bibitem[Dong \latin{et~al.}(2020)Dong, Du, and Gardner]{Dong:2020aa}
Dong,~E.; Du,~H.; Gardner,~L. {An interactive web-based dashboard to track
  COVID-19 in real time}. \emph{The Lancet Infectious Diseases} \textbf{2020},
  \emph{3099}, 19--20\relax
\mciteBstWouldAddEndPuncttrue
\mciteSetBstMidEndSepPunct{\mcitedefaultmidpunct}
{\mcitedefaultendpunct}{\mcitedefaultseppunct}\relax
\EndOfBibitem
\end{mcitethebibliography}

\providecommand{\latin}[1]{#1}
\makeatletter
\providecommand{\doi}
  {\begingroup\let\do\@makeother\dospecials
  \catcode`\{=1 \catcode`\}=2 \doi@aux}
\providecommand{\doi@aux}[1]{\endgroup\texttt{#1}}
\makeatother
\providecommand*\mcitethebibliography{\thebibliography}
\csname @ifundefined\endcsname{endmcitethebibliography}
  {\let\endmcitethebibliography\endthebibliography}{}

%
%
\end{document}